# Particle Sources

*D.C. Faircloth*
Rutherford Appleton Laboratory, Chilton, Oxfordshire, UK

**Abstract**
This paper outlines the many ways that the initial beam can be made for particle accelerators. Brief introductions to plasma physics and beam formation are given. Thermionic and photo emission electron guns, with both DC and Radio Frequency (RF) acceleration are outlined. Positive ion sources for producing $H^+$ ions and multiply charged heavy ions are covered. Hot cathode filament sources and cold cathode sources are explored. RF discharge sources (inductively coupled, microwave and ECR) are discussed, as are laser, vacuum arc, and electron beam sources. The physical principles of negative ion production are outlined and different types of negative ion source technologies are described. Polarised particle sources are mentioned briefly. Source choice is summarised and the general practicalities of source operation and development are discussed.

## 1    Introduction

There is a vast zoo of different particle sources out there with many varied applications. Some need to produce ions from tiny samples so they can be analysed and measured. Some need to produce electron beams for imaging, or for welding, or for producing coherent synchrotron light, or for Free Electron Lasers (FEL). Some need to produce ions for industrial processes such as coating, etching, implanting and isotope separation. Some will go into space to provide thrust for satellites. Fusion demands ion sources that generate huge currents of hundreds of amps with beam cross sections measured square meters. Discovery machines demand ever higher beam currents in ever smaller cross sectional areas. Radioactive rare isotope ion sources and neutron sources require an entire accelerator facility to drive them.

This paper will concentrate on particle sources for accelerators and so will focus on sources that produce a beam with a small cross section and low divergence (emittance). Particle sources are a critical component of all particle accelerators. They create the initial beam that is accelerated by the rest of the machine, so they define the overall performance of the particle accelerator. Electron sources are generally called electron guns or just guns. Ion sources are generally called sources.

All particle sources consist of two parts: A **particle generator** and an **extraction system**.

The **particle generator** must be able to provide enough of the correct particles to the extraction system. There are numerous ways of generating particles. The most common is to create a plasma with an electrical discharge. Electron guns generally employ thermionic and photoemission from surfaces. Particles can be generated by sputtering from surfaces or by interactions with other particles in gases. Lasers can be used for ionisation or excitation. Particle sources often employ combinations of processes. The particle generator must be stable for long enough to extract a beam for whatever the application requires.

The **extraction system** must be able to produce a beam of the correct shape and emittance for the next stage of the accelerator by extracting the correct particles from the particle generator and removing unwanted ions, electrons or neutral particles. For particles with very similar charge-to-mass ratios the removal of unwanted ions might take place later downstream in bending dipoles or accelerating structures. The extraction system uses electric fields to accelerate the particles and magnetic fields to separate them. The electric fields are shaped by carefully designed electrodes with voltages applied to them. The magnetic fields are produced by currents flowing in coil windings or by permanent magnets, the magnetic fields can be shaped and confined by magnetic pole pieces and yokes. Apertures and dumps are positioned to select the correct species and safely get rid of unwanted species. Sometimes electrodes in the beam optical system that make up the extraction system are also used to supress particles of opposite charge back-streaming into the ion source.

## 2 Plasma and Physics Concepts

### 2.1 Introduction

A common way of making ions is by creating a plasma. Plasma is the fourth state of matter: if you keep heating a solid, liquid or a gas it will eventually enter the plasma state. However, entering the plasma state is not a distinct phase change like the melting and boiling transitions, it is a continuous transition as more and more of the atoms and molecules become ionised. 'Percentage ionisation' is a measure of how ionised the plasma is, i.e. what proportion of the atoms and molecules have been ionised.

Plasma contains approximately equal proportions of positive and negative charges in the form of positive ions, negative ions and electrons. All the composite particles (everything other than electrons and protons) can also be put into excited states: electrons in atoms and molecules can be elevated into higher orbitals; molecules can be made to bend and vibrate in various modes and states; molecules can be made to spin in different rotational states.

Charged particles are always electrostatically attracting or repelling each other. Collisions can cause excitation, ionisation, neutralisation or other disassociation/combination reactions. Photons are constantly being emitted and absorbed in these reactions. The physics of plasmas can be extremely complex with many fascinating emergent behaviours caused by the radically different masses of electrons and ions, and the myriad of reactions that occur at rates highly dependent on the velocity of the particles.

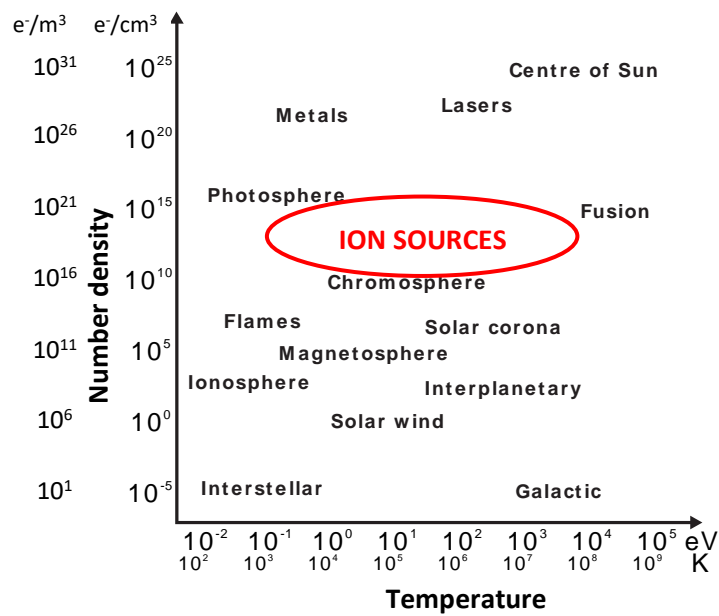

**Fig. 1:** Different types of plasma.

What follows are some of the key plasma physics properties and concepts that relate to particle sources.

## 2.2 Basic Plasma Properties

### 2.2.1 *Range of plasmas*

Plasmas exist in nature over a huge range of densities and temperatures. Some examples are shown in Fig. 1. Note that although flames are shown, they are very weakly ionised (<0.0001% ionisation).

### 2.2.2 *Number Density, n*

The most basic parameter is the particle number density of each of the constituents in the plasma. It is usually written as *n* with a subscript to represent the type of particle e.g.:
- $n_e$ = number density of electrons
- $n_i$ = number density of ions
- $n_n$ = number density of neutrals.

It is expressed in terms of number of particles per cm³ or per m³ of volume. For comparison, atmospheric air has a number density of around $10^{25}$ m$^{-3}$.

### 2.2.3 *Temperature, T*

Temperature is a measure of how much particles are moving, their kinetic energy. In a plasma (where the particles are free), a particles velocity (and mass) defines its kinetic energy.
The Boltzmann constant gives 11600K = 1 eV. The temperature is usually expressed in eV.

- $T_e$ = temperature of electrons
- $T_i$ = temperature of ions
- $T_n$ = temperature of neutrals.

The temperature of the electrons, ions and neutrals can be different to each other. Different types of plasma produced in different ion sources can have very different ion and electron temperatures. For example: Electron Cyclotron Resonance (ECR) ion sources (Section 5.6.4), where the plasma is heated by repeatedly accelerating the electrons, can have $T_e > 1$ keV and $T_i < 1$ eV.

### 2.2.4 *Temperature Distribution*

Obviously not all the particles of the same type will have the same temperature. When talking about the overall temperature of a plasma (rather than the temperature of a single particle), the numbers $T_e$, $T_i$ and $T_n$ are statistical fits to Maxwellian distributions. If the plasma is in thermal equilibrium it will obey Maxwell-Boltzmann statistics and a temperature can be defined.

Using the standard equation for kinetic energy the r.m.s. speeds can be calculated to be:

$$velocity\ of\ electrons, \bar{v}_e = 67\sqrt{T_e} \tag{1}$$

$$velocity\ of\ ions, \bar{v}_i = 1.57\sqrt{\frac{T_i}{A}}\ , \tag{2}$$

where *A* is the ion mass in atomic mass units and $T_e$ and $T_i$ are in K.

If the plasma is not in thermal equilibrium (i.e. in a transient or pulsed state) then a single temperature value cannot reasonably be defined, instead an Energy Distribution Function (EDF) showing a histogram of the number of particles with each energy is a better description of the thermal properties of the plasma.

Often the plasma is in a magnetic field. The ions and electrons will spiral around the magnetic field lines and slowly move along them. It the plasma heating technique favours a particular direction, by virtue of the applied field direction, the particle velocities (temperatures) will not be the same in all directions. $T_{i\parallel}$ is the ion temperature parallel to the magnetic field and $T_{i\perp}$ is the ion temperature perpendicular to the magnetic field.

### 2.2.5 Charge State, q

The charge state of an ion indicates how many electrons have been removed from it. Singly charged ions have a charge state $q = +1$. Not all ions will be singly charged, some ions will be multiply ionised (e.g. $Pb^{3+}$ which has a charge state $q = +3$). Some ions will be negatively charged (e.g. $H^-$ which has a charge state $q = -1$). Only those elements (or molecules) which have a positive electron affinity can form negative ions. The number densities, $n$, of each charge state can be very different.

### 2.2.6 Quasi Neutrality

Plasma is generally charge neutral, so all the charge states of all the ions adds up to the same number as the number of electrons.

$$\sum q_i \, n_i = n_e \tag{3}$$

### 2.2.7 Ionisation Energy

The ionisation energy is the energy required to remove an electron from an atom, or to put it another way, it is the energy that must be gained by an electron for it to escape from its atom. The larger the atom, the easier it is to remove the outermost electron. The second electron is always harder to remove, the third even harder and so on. In most ion sources the energy for ionisation comes from electrons impacting on neutral atoms or molecules in electrical discharges. The electrons get their energy by being accelerated in the electric field applied to the discharge.

### 2.2.8 Percentage Ionisation

As mentioned previously, percentage ionisation is a measure of how ionised the gas is, i.e. what proportion of the atoms are ionised.

$$percentage\ ionisation = 100 \times \frac{n_i}{n_i + n_n} \tag{4}$$

When the percentage is above 10% the plasma is 'highly ionised' and its behaviour becomes dominated by plasma physics interactions. At less than 1% ionisation the plasma is 'weakly ionised' and the physics of neutral particle interactions play an important role in the plasma behaviour.

The transition into the plasma state is not a sudden phase change it is a continuum as more and more atoms or molecules become ionised. The energy to ionise comes from the kinetic (thermal) energy of the impacting particles. The ionising particles are predominantly electrons, but if the gas is hot enough the kinetic energy of the atoms and molecules themselves can cause ionisation when they collide with each other. This self-ionisation occurs above 3000K and is described by the Saha ionization equation.

### 2.2.9 Mean Free Path vs Relaxation time

Collisions between particles in a plasma are fundamentally different from collisions in a neutral gas. The ions in a plasma interact by the Coulomb force: they can be attracted or repelled from a great distance. As an ion moves in a plasma its direction is gradually changed as it feels the electric fields around neighbouring charged particles. Whereas in a neutral gas, the particles only interact when they get so close to each other that they bounce off each other's outer electron orbitals. In a neutral gas the average distance the particles travel in a straight line before bouncing off another particle is referred to as the mean free path. In a plasma, the mean free path concept does not work because the ions and electrons are always interacting with each other by their electric fields. Instead a concept called 'relaxation time' is invoked: this is the time it takes for an ion to change direction by 90°. The relaxation time $\tau_\theta$ can also be described as 'the average 90° deflection time'. In a plasma, different particle species can have different relaxation times.

*2.2.10 Reaction Cross Sections*

When two particles collide a reaction might happen if the particles have the correct energy. The type of reaction that occurs (ionisation, excitation, detachment, attachment, etc.) depends on the type of particles involved and their energy (which is a combination of their speed and excitation). The likelihood (probability) of a reaction occurring is strongly dependant on the energy of the particles and it is expressed in terms of a cross sectional area. This intuitively makes sense: objects with a larger area are more likely to be hit.

Cross sections vary with energy. For an ionisation reaction, if an electron is going too slowly, the target particle (an atom or molecule) is less likely to be ionised, and if the electron's kinetic energy is below the ionisation energy of the particle, the reaction cross section rapidly decreases to zero. If an election is going too fast it is less likely to interact with the other particle, so the cross section becomes smaller. For elastic scattering of electrons the cross section does not really have a minimum; two electrons will always have a Coulomb repulsion between them. The variation of cross section with energy for a reaction may have many ups-and-downs and the cross section might not drop to zero depending on the type and complexity of the physics involved in the particle interaction.

*2.2.11 Reaction Rates and Collison Rates*

Cross section data can be combined with particle number densities and energy distributions to obtain reaction rates for the different reactions that occur in a plasma (or conversely cross sections can be obtained from reaction rates). As a plasma gets denser the reaction rates increase because the particles are colliding more often (the collision rate increases). As a plasma gets hotter the reaction rates increase. This is because the cross sections must be weighted by the particle velocities to obtain the rate at which particles encounter each other. The reaction rate is defined as the product of the particle number densities and the average of the energy-dependent cross section weighted by the relative velocity of the colliding particles. For electron impact collisions the motion of the neutral particle or ion can be neglected, thus the relative velocity of the colliding particles is the electron velocity.

## 2.3 Surfaces

*2.3.1 Overview*

Surfaces play various important roles in particle sources. Surfaces can be used as a source of electrons. The electrons can be accelerated away from the surface and they can be used to produce negative ions at the surface. Plasmas are bounded by surfaces, the ways particles interact with surfaces can be important to the performance of the particle source. Plasma interacting with surfaces including insulator surfaces can produce molecules with a wide range of rotational and vibrational excitations. Excited states can be used to drive the reactions that produce the desired particle species. Ions in a plasma can knock atoms off surfaces by a process called sputtering. This process moves material around the particle source and is an important factor that limits source lifetime. Sputtering can be used as a way of producing ion beams from solid and hard-to-vaporise materials.

Particles leaving the surface can be of a different charge state from the incoming particle, or be excited, or be in a molecule, or some combination of all three. Complex interactions can take place at surfaces. The surface itself may also be altered. The condition of the surface can be critical to the operation of the particle source.

When a particle interacts with a surface many complex and competing processes can occur:
− Reflection
− Adsorption
− Sputtering
− Desorption

- Recombination
- Dissociation
- Ionisation
- Secondary electron emission
- Photo emission
- Excitation.

### *2.3.2  Work Function*

In any metal, there are one or two electrons per atom that are free to move from atom to atom. This is sometimes collectively referred to as a 'sea of electrons'. Rather than being uniform, the electron velocities in this sea follow a statistical distribution. At absolute zero the electrons are all in their lowest energy state and follow Fermi-Dirac statistics. Figure 2 shows a histogram of the electron energy distribution in a metal. The work function is the amount of energy required for an electron to exit the metal without being pulled back in by the positive ion lattice. The energy of an electron that is able to leave the surface is defined as $E_0$ or the 'vacuum level' because this is the energy required for an electron to leave the surface and enter a vacuum and then theoretically drift away to infinity. The work function is the difference between the 'vacuum level' and the 'Fermi level'. The work function is a characteristic of the material and for most metals is of the order of several eV.

Continuing the sea of electrons analogy: at 0 K the electron sea is calm and the height of the harbour wall is the work function.

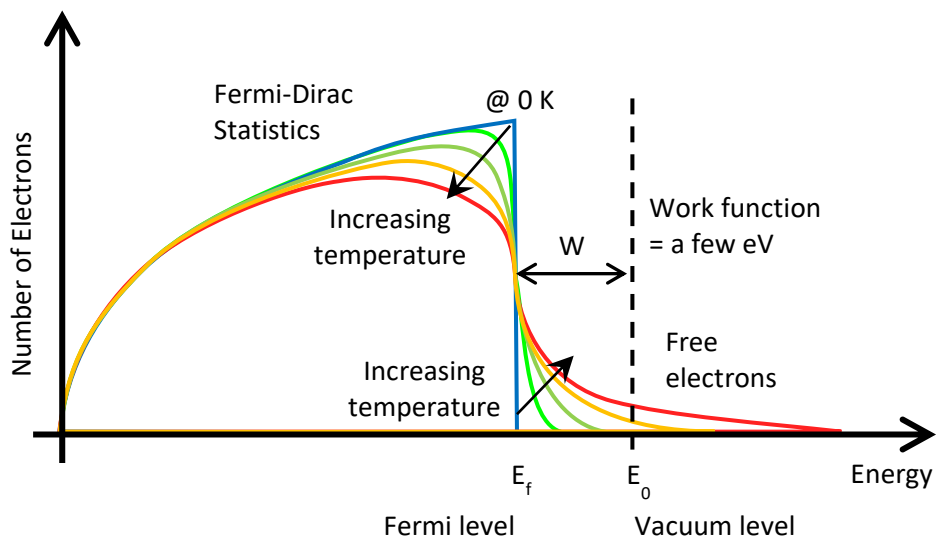

**Fig. 2:** Electron energy distribution function in a metal at different temperatures with the work function shown.

### *2.3.3  Thermionic emission*

Thermionic emission is the heat-induced flow of electrons from a surface. Thermionic emission has been studied by numerous researchers. If the temperature of any material is increased (as shown in Fig. 2.) the electrons will start to have a range of energies around the Fermi level and eventually some of the higher energy electrons will be fast enough to escape the 'sea of electrons' and become free electrons on the surface of the material (as shown by the orange and red lines reaching over the dashed line in Fig. 2). The fastest electrons have enough thermal energy to overcome the work function of the material. The sea of electrons is rough enough to start splashing over the harbour wall.

Thermionic currents can be increased by increasing the temperature of the metal or by decreasing the work function. The often-desired goal of a low work function surface can be achieved by applying various oxide coatings to the surface.

In 1901 Owen Richardson found that the current from a heated wire cathode varied exponentially with temperature. He later proposed this equation:

$$J = A_G T^2 e^{\frac{-W}{kT}}, \qquad (5)$$

where:
- $J$ is the electron current density on the surface of the cathode
- $W$ is the cathode work function
- $T$ is the temperature of the cathode.

$A_G$ is given by:

$$A_G = \lambda_R A_0, \qquad (6)$$

where $\lambda_R$ is a material-specific correction factor that is typically of order 0.5, and $A_0$ is a universal constant given by:

$$A_0 = \frac{4\pi m_e k^2 q_e}{h^3} = 1.20173 \times 10^6 \text{ Am}^{-2}\text{K}^{-2}, \qquad (7)$$

where $m_e$ and $q_e$ are the mass and charge of an electron, and $h$ is Planck's constant.

### 2.3.4 Field Enhanced Thermionic Emission

The electron emitting surface is almost always in an electric field (because the electrons are pulled away by the extraction field). Depending on the direction of the electric field (either into or out of the surface), it can either enhance or reduce the thermionic current by effectively changing the work function by an amount ΔW.

Equation (5) can be modified to:

$$J = A_G T^2 e^{\frac{-(W-\Delta W)}{kT}} \qquad (8)$$

ΔW is given by:

$$\Delta W = \sqrt{\frac{q_e^3 F}{4\pi\epsilon_0}}, \qquad (9)$$

where: $F$ is the electric field strength normal to the surface.

As the field is increased, more of the fast electrons are allowed to escape the metal. The way the field controls the current is called the Schottky effect, named after the German physicist Walter Schottky.

Changing the electric field at the surface is analogous to changing the height of the harbour wall that bounds the sea of electrons- lower the harbour wall and fast electrons will flood over, raise the harbour wall and the fast electrons can no longer escape. This principle is used in electronic components in the very fast acting Schottky diode, which for obvious reasons is also called a 'hot-carrier' diode.

### 2.3.5 Field Emission

Equation (8) for field enhanced thermionic emission is valid for fields up to about 0.1 GVm$^{-1}$. At very high electric fields, quantum tunnelling of electrons becomes significant. Field emitted electrons do not come from the hot end of the electron distribution shown in Fig. 2. Instead, electrons with energies near the Fermi level, quantum tunnel through the reduced work function potential barrier. Field emission is used in electron guns for imaging and for other applications, but rarely for particle accelerators. Field emission is still important in particle accelerators because of the role it plays in high voltage breakdown of vacuum insulation systems. Field emission requires very high fields which are usually produced around geometric field enhancements, i.e. electrode points or triple junctions.

*2.3.6    Photoemission*

Photoemission was first observed by Heinrich Hertz in 1887 where he showed high voltage gaps would breakdown when they were illuminated by Ultra Violet (UV) light. The mystery of why longer wavelength light would not produce the same effect no matter how intense the light, was solved by Einstein with his theoretical explanation of Photoemission in 1905. When photons hit a surface, their electromagnetic wave packet transfers energy to electrons in the surface. If there is enough energy in a photon (its wavelength is short enough) it can impart enough energy to an electron to allow it to escape the positive ion lattice (it can overcome the material's work function). If there is not enough energy transferred from the photon for the electron to escape, the electron will rapidly thermalize (share its energy with its neighbours), no matter how many photons hit the surface, there will be no electrons with enough energy to overcome the surface work function.

The Quantum Efficiency (QE) of a surface is the number of electrons produced per incident photon. Copper is often used as a photocathode even with its very low QE because it can withstand very high power densities due to its high thermal conductivity. The QE of a surface significantly varies with the wavelength of the incoming photon.

## 2.4   Magnetic Confinement

Charged particles will rotate around magnetic field lines, this means they tend to travel along magnetic field lines, by spiralling along them. This effect can be exploited to confine plasmas. A dipole field will confine ions in the direction of the magnetic field this can be used to confine electrons between two parallel cathodes (Penning discharge). Similarly, a solenoidal field will keep charged particles confined axially. Particle transport perpendicular to the field occurs through collisions.

A multicusp field is comprised of an array of alternating north and south poles of permanent magnets. This arrangement is used around the edge of a plasma chamber to confine both electrons and ions, preventing the plasma hitting the walls of the chamber (except at the cusps). A multicusp arrangement of magnets is repeated around the outside of a cylindrical plasma chamber or on a flat surface of a plasma chamber. The multicusp arrangement can be used on very large plasma chambers. Multicusp plasma confinement was originally developed in the 1970s for fusion research, with dimensions of the order of 1 m these sources truly are giant buckets of plasma. This is why ion sources employing multicusp confinement are sometimes referred to as 'bucket' sources.

A multipole field arrangement is similar to the multicusp, but only applied around the outside of a cylindrical plasma chamber and named after the number of poles, e.g. hexapole is commonly used in electron cyclotron resonance sources (see Section 5.6.4).

## 2.5   Debye Length

Named after the Dutch scientist Peter Debye, the Debye length, $\lambda_D$, is the distance over which the free electrons redistribute themselves to screen out electric fields in plasma. In one Debye length the field drops to 1/e of the unscreened field. This screening process occurs because the light mobile electrons are repelled from each other whilst being pulled by neighbouring heavy (low mobility) positive ions, thus the electrons will always distribute themselves between the ions. Their electric fields counteract the fields of the ions creating a screening effect. The Debye length not only limits the influential range that particles' electric fields have on each other but it also limits how far electric fields produced by voltages applied to electrodes can penetrate into the plasma. The Debye length effect is what makes the plasma quasi-neutral over long distances.

The higher the electron density the more effective the screening, thus the shorter this screening (Debye) length will be.

The Debye length is given by:

$$\lambda_D = \sqrt{\frac{\epsilon_0 k T_e}{n_e q_e^2}}, \tag{10}$$

where:

- $\lambda_D$ is the Debye length, which is of the order of 0.1 – 1 mm for ion source plasmas
- $\epsilon_0$ is the permittivity of free space
- $k$ is the Boltzmann constant
- $q_e$ is the charge of an electron
- $T_e$ is the temperatures of the electrons
- $n_e$ is the density of electrons.

## 2.6 Plasma Sheath

The screening effect of the plasma creates a phenomenon called a plasma sheath around any plasma-surface boundary. The plasma sheath is also called the Debye sheath or electrostatic sheath. The potential of the bulk of a plasma is close to the anode voltage with the transition to other electrode voltages occurring in the plasma sheaths that surround them. For DC plasmas the largest potential drop is across the sheath near the cathode electrode. The cathode sheath has a greater density of positive ions, and hence an overall excess positive charge. It balances an opposite negative charge on the cathode with which it is in contact. The plasma sheath is several Debye lengths thick.

A related phenomenon is the double layer which can occur in current carrying plasmas.

## 2.7 Electrical Discharges

### 2.7.1 Overview

The vast majority of plasmas used in ion sources are generated by an electrical discharge. The driving field applied to a discharge is often an electrical field, however a magnetic field can also be applied in the case of inductively coupled discharges (but it is still ultimately the induced time-varying-electric-field that actually accelerates the electrons). Discharges produced by time varying fields are more difficult to analyse, so it is best to start with an explanation of a DC electric field discharge between two electrodes. The general current and voltage characteristics of such a discharge are summarised in Fig. 3. The exact shape of the curve depends on the type of gas, pressure, electrode geometry, electrode temperatures, electrode materials, electrode surface roughness and any magnetic fields present.

### 2.7.2 Dark Discharge

At low voltages the current between two electrodes is very small, but it slowly increases as the voltage between the electrodes increases as shown in the bottom left corner of Fig. 3. This tiny current comes from ions and electrons that have been produced by background ionisation. These are swept out of the gap by the electric field between the electrodes that is created by the applied voltage. There are only enough charge carriers produced by background radiation and cosmic rays for a few nA of current, so the current quickly saturates. The voltage can then be increased with no increase in current. The ions and electrons are pulled towards the electrodes through the gas molecules interacting with them as they go.

*2.7.3    Townsend Breakdown*

Eventually the applied electric field is high enough to accelerate the electrons to the ionisation energy of the gas. At this point the discharge current rapidly increases as shown in bottom right corner of Fig. 3. The electrons ionise the neutral atoms and molecules, producing more electrons. These additional electrons are accelerated to ionise even more atoms producing even more free electrons in an avalanche breakdown process known as Townsend breakdown. This runaway process means the voltage needed to sustain the discharge drops significantly.

What happens next depends on the electrode geometry, gas pressure and the power supply that is driving the discharge. If the electrodes are very pointy a corona discharge might form. The electric field around a point is very high, but decreases rapidly away from the point. So, discharge avalanches that start in the local field enhancement around the point propagate into a rapidly decreasing field and go out. Corona discharges are spatially limited around points.

If the avalanche process bridges the gap the impedance between the electrodes decreases rapidly. The power supply must be designed to control the discharge otherwise it will just trip-out or damage itself. Depending on the gas pressure and electrode geometry the discharge might go straight to the arc discharge regime. If the pressure is low enough (mBar range) a glow discharge might form. If the gap is long enough transient filamentary discharges such as streamers and leaders might form.

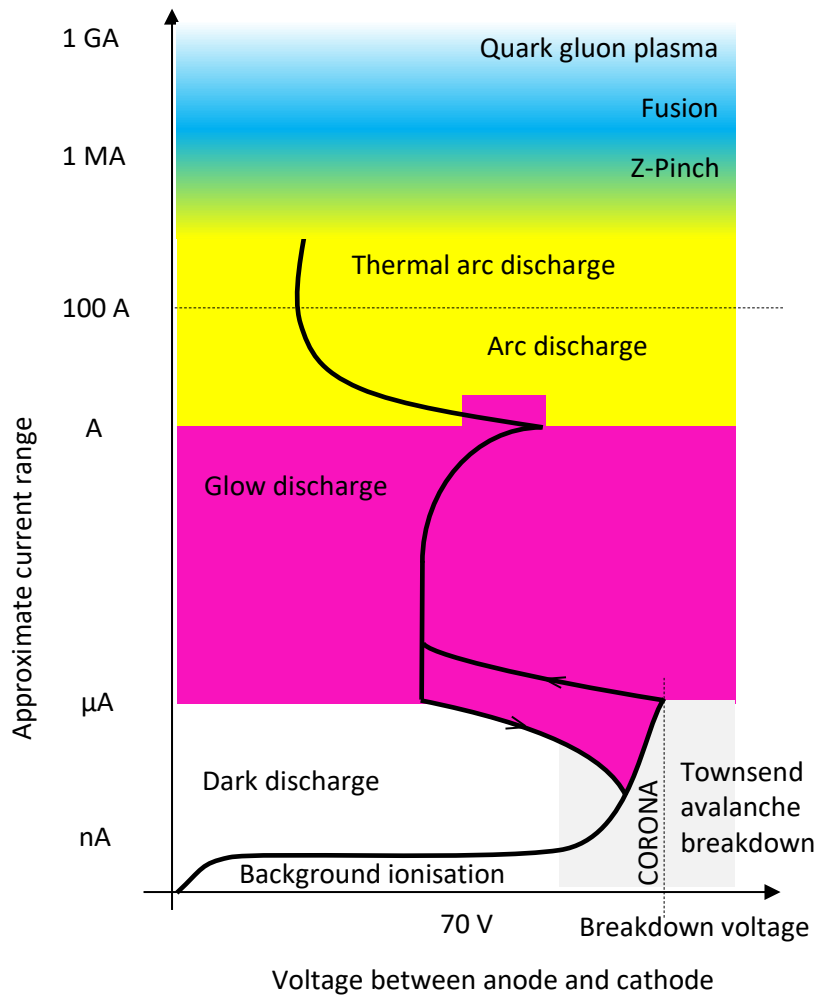

**Fig. 3:** The current voltage characteristics of electrical discharges.

*2.7.4 Glow Discharge*

The glow discharge is so called because it emits a significant amount of light. Most of the photons that make up this light are produced when atoms that have had their orbital electrons excited by electron bombardment, relax back to their ground states. More generally photons are produced in any event that needs to release energy. For example: when ions recombine with the free electrons and when vibrationally excited molecules relax.

A glow discharge is self-sustaining because the accelerated positive ions impact on the cathode, producing more electrons in a process called secondary emission. This is why in there is a hysteresis in the current voltage curve at the glow to dark discharge transition.

The current in a glow discharge can be increased with very little increase in discharge voltage. The plasma distributes itself around the cathode surface as the current increases. Eventually the current reaches a point where the cathode surface is completely covered with plasma and the only way to increase the current further is to increase the current density at the cathode. This causes the plasma voltage near the cathode to rise.

*2.7.5 Arc Discharge*

The increased current density leads to cathode heating and eventually the cathode surface reaches a temperature where it starts to thermionically emit electrons and the discharge moves into the arc regime with a negative current voltage characteristic where the voltage across the discharge drops as the current rises.

The name 'arc discharge' is historical. Electrical arcs are typically free, unbounded discharges between two electrodes. However, any source plasma with a negative current voltage gradient is often called an 'arc' or it is said to be running in an 'arc regime' or a 'negative impedance' regime.

The current increases in an arc until the plasma is almost completely ionised (it has reached a very high ionisation percentage with few neutral particles left). Eventually the current density in the plasma reaches a point where the ions have the same average energy (temperature) as the electrons. They have reached thermal equilibrium. The discharge enters the thermal arc regime where the discharge voltage rises as the current increases.

As the discharge current increases further the magnetic field created by the discharge current itself starts to have a significant effect and the Lorenz force squeezes the discharge together in a phenomenon called Z-Pinch making the plasma even denser. At extremely high discharge currents the plasma becomes so dense that atomic nuclei start fusing together and eventually the nuclei are compressed to a quark-gluon plasma. These extreme discharges are highly unstable and can only be achieved in incredibly short transient pulses.

## 2.8 Importance of the Power Supply

As already mentioned in Section 2.7.3 the power supply must be designed operate with the desired type of discharge. The discharge current and voltage obtained will be where the power supply load curve intersects the characteristic shown in Fig. 3. The gradient of the discharge characteristic at the intersection point determines if the discharge is stable or not. If operating in the arc regime a series resistance allows a power supply to successfully regulate the current. The power supply needs to produce a reasonably high 'strike potential' to start the discharge, then be able to operate at a lower voltage, this is usually at the cost of the electrical efficiency of the power supply (the volts that are not dropped across the discharge have to be dropped across internal devices that dissipate power).

## 2.9 Paschen Curve

The breakdown voltage of any gas between two flat electrodes depends primarily on the electron mean free path and the distance between the electrodes. The mean free path is the average distance particles travel before hitting other particles. It is directly related to pressure. Figure 4 shows how the breakdown voltage of hydrogen varies with the product of pressure, $p$ and distance, $d$ between electrodes. This was first stated in 1889 by Friedrich Paschen [1].

At very low pressures, the mean free path between collisions is longer than the distance between the electrodes. So, although the electrons can be accelerated to ionising energies, they are unlikely to hit anything other than the anode. This means that the breakdown voltage is very high at very low pressures. At very high pressures the mean free path is very short. This means that the electrons never have enough time to be accelerated before hitting another particle. Between these two extremes exists a minimum which occurs at about 1 mBar cm for all gasses. The exact position of this minimum depends on the type of gas and the electrode properties.

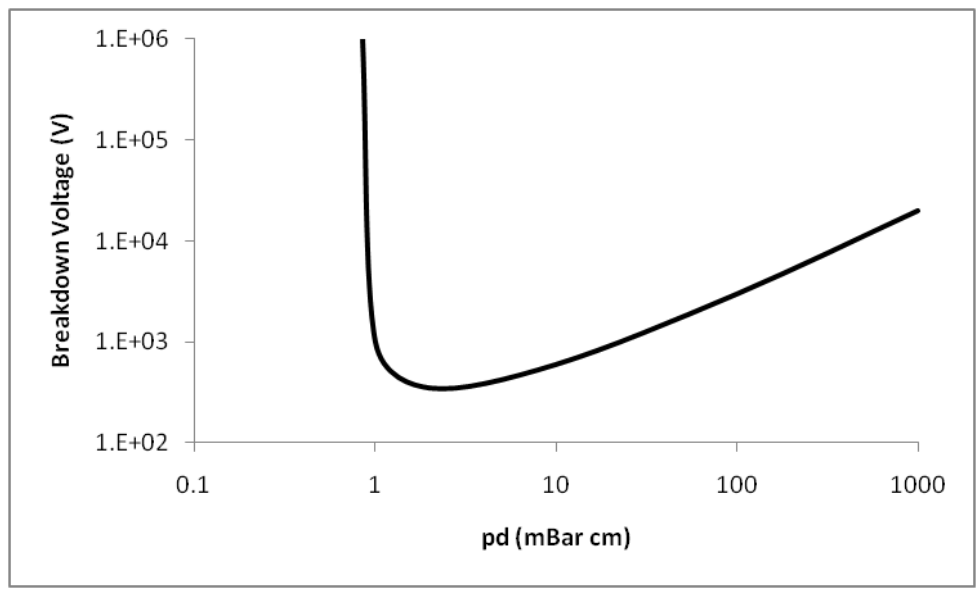

**Fig. 4:** The Paschen curve for hydrogen.

## 3 Extraction

### 3.1 Introduction
The purpose of the extraction system is to produce a beam from the plasma generator and deliver it to the next acceleration stage. The basics of extraction are very simple: apply a high voltage between a particle emitting surface and an extraction electrode with a hole in it. The extraction electrode can also be called the acceleration electrode or the ground electrode (because the particle source is biased at a high voltage and the extracted beam is accelerated to ground).

### 3.2 Meniscus Emitting Surface
In plasma based ion sources the ion emitting surface is the edge of the plasma itself. At the extraction region the plasma is bounded by an electrode with a hole in it. This electrode is variously called the outlet electrode, aperture electrode, or plasma electrode and it is often held at (or very close to) the same potential as the plasma anode. The edge of the plasma sits across the hole and is called the plasma meniscus. It is the boundary layer between the discharge and the beam. The meniscus is not actually a discrete boundary like the meniscus on the surface of a liquid, it is a few Debye lengths thick, however it is useful to think of it as a surface when qualitatively considering the particle trajectories.

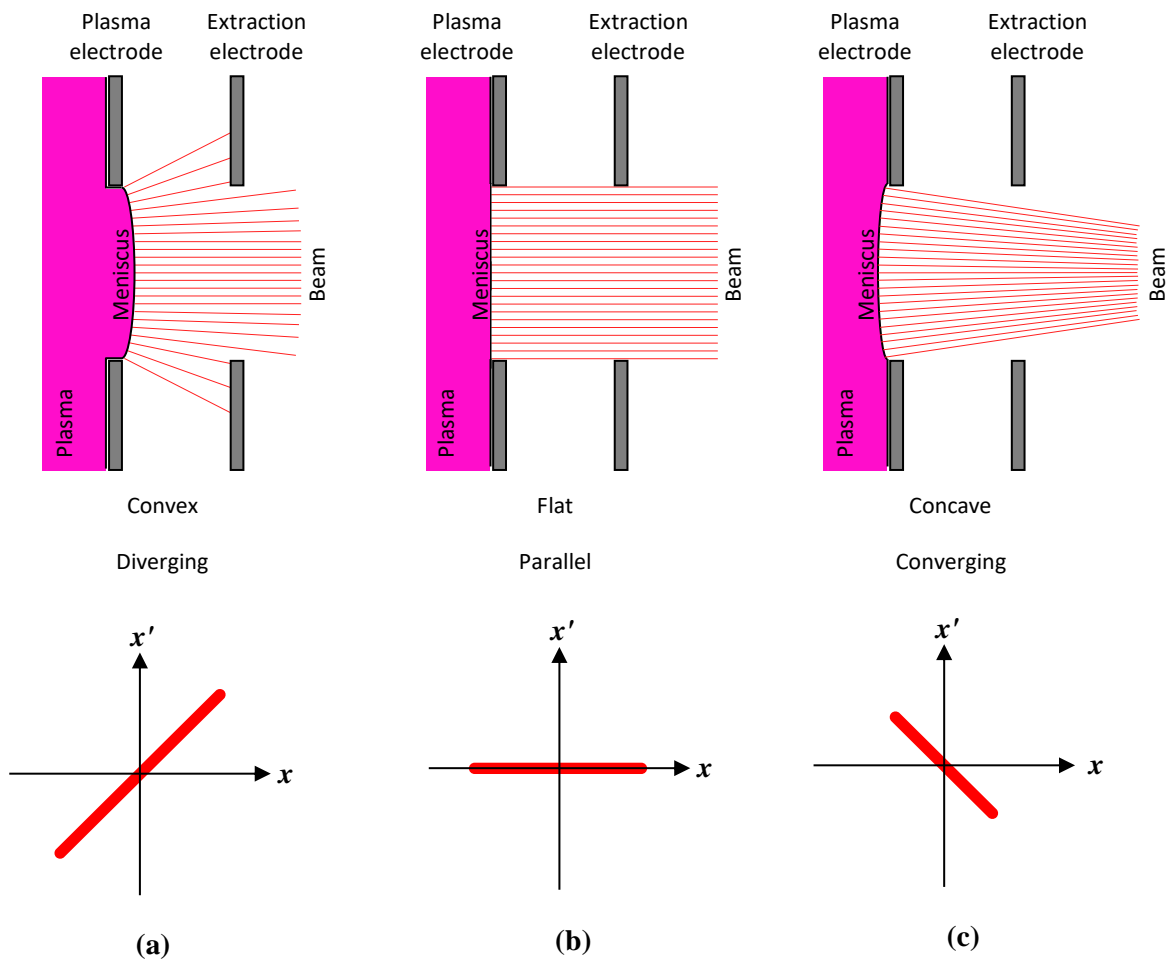

**Fig. 5:** Three different plasma meniscus shapes and their corresponding emittance phase space plots.

The shape of the meniscus depends on the local electric field and the local plasma densities. Figure 5 shows how the plasma meniscus can change from being convex to concave as the plasma density decreases. The trajectories of the particles depend on the meniscus shape, so it is important to run the ion source with operating conditions that provide an optimum meniscus shape. This is called the 'matched case' (the beam is matched to the ion optics) and it is found by varying the plasma density and extraction potential until the beam is well transported. It is important to point out that the diagrams in Fig. 5 do not include space charge effects that cause the beam to diverge (see Section 3.7).

### 3.3 Solid Emitting Surface

In caesiated negative ion sources like the surface converter source (see Section 6.4.3) the negative ions are produced on an electrically biased electrode surface inside a plasma chamber. This has an advantage over meniscus emission in that the exact shape of the surface can be precisely defined. The emission surfaces is concave with a radius of curvature approximately centred on the exit aperture of the plasma chamber. This causes the ions produced on the emission surface to be focussed towards the exit aperture, resulting in a stable low emittance beam.

### 3.4 Emittance

It is essential that the beam produced by the ion source has a low divergence angle. This allows the beam to be transported and accelerated easily by the rest of the machine without losing any beam.

In particle accelerators a way of specifying the divergence of a beam is emittance ($\epsilon$). Emittance is a measurement of how large a beam is and how much it is diverging. It is measured in mm.mRads and is the product of beam size and divergence angle. Often emittance is normalised to beam energy because a beam that has had its longitudinal velocity increased by acceleration will still have the same transverse velocity, thus its divergence angle will be reduced. Emittance is normalised to allow emittances to be compared at different energies in different accelerators.

For any beam two emittances are given: horizontal and vertical. These can be just single values ($\epsilon_x$ and $\epsilon_y$) or complete phase space diagrams. Phase space diagrams are plots of divergence angle versus transverse position. Fig. 5 gives examples of phase space plots for divergent, parallel and convergent beams. If the horizontal axis of the emittance phase space diagram is *x* then the vertical axis is usually given as *x'* (where *x'* is $v_x$ or $\frac{dx}{dt}$), this is because the transverse velocity (or momentum) of the particles defines the divergence angle.

The emittance of a beam is proportional to the area enclosed by its phase space plot. For real beams this statement needs clarification. Real beams have halos- outlying particles that have much larger divergence angles and positions than the core of the beam. They are created when some particles at the edge of the beam experience fringe fields or other non-uniformities that cause them to separate further from the core of the beam. The halo particles are in the minority, most of the particles are in the dense core of the beam. If these outlying particles are included in the total phase space area calculation the beam will have a huge emittance. To get round this emittances are either quoted as the 95% emittance or r.m.s. emittance. The 95% emittance is the area that encloses 95% of all the particles. The r.m.s. emittance is calculated from the r.m.s. values of all the positions and angle measurements. Both methods give a fair measurement of the overall beam divergence without it being affected by halo particles.

The r.m.s. emittance is defined as:
$$\epsilon_{x_{rms}} = \sqrt{\langle x'^2 \rangle \langle x^2 \rangle - \langle xx' \rangle^2} \,, \tag{11}$$
where the triangular brackets are the expectation values.

The use of the $\pi$ symbol in emittance units should also be mentioned. Often emittances are expressed in units of $\pi$.mm.mRad. This is because the area in emittance phase space plots is often drawn as an ellipse and the $\pi$ relates to calculating the area of an ellipse from a product of the length of its two semi-axes.

### 3.5 Energy Spread

In a beam not all the particles have the same energy. The energy distribution of the particles is a measurement of the range of different particle velocities in the beam. It is effectively the longitudinal emittance of the beam measured in eV. It is more commonly defined as the full-width-half-maximum of the energy distribution in eV. It is sometimes also referred to as the momentum spread. In emittance phase space plots a proportion of the 'thickness' of the phase space distribution is caused by energy spread.

The energy spread causes the beam to spread out in time, so this limits the minimum pulse length achievable. The origin and size of the energy spread is different for different types of source. It can be caused by: variations in potentials or temperatures on the plasma production surface; oscillations in the plasma; or unstable extraction voltages. The energy spread can range from less than 1 eV to as much as 100 eV.

Energy spread is important because it will produce transverse emittance growth as the beam passes though magnets and accelerating gaps. The beam emittance can be transferred between longitudinal and vertical directions and vice versa.

### 3.6 Brightness

The brightness of a beam is another key beam parameter. It is the beam current, $I$ divided by the emittances:

$$B = \frac{I}{\epsilon_x \epsilon_y} . \tag{12}$$

Unfortunately, there are several ways to define brightness, they all have the same basic form as Eq. (12) but they have each have different scale factors based on multiples of $\pi$ and 2. The reader should take caution when comparing brightness from different authors. Also, the emittances can be normalised to energy, which gives an emittance normalised brightness.

### 3.7 Space Charge

Space charge effects are critical in ion source design. For high-brightness low-energy beams, electrostatic forces are a key factor. The beam will blow-up under its own space charge so it is critical to get the beam energy up to at least 10 keV as fast as possible to minimise the effect which is worse at low energies.

A phenomenon known as 'space charge compensation' or 'space charge neutralisation' is essential for high current ion sources. The pressure in the vacuum vessel directly after extraction will be higher than in the rest of the accelerator because of gas loading from the ion source itself. The beam ionises the background gas as it passes through it. If the beam is positive it repels the positive background ions and draws in the negative ions and electrons. If the beam is negative it repels the negative background ions and electrons and draws in the positive ions. The effect is to neutralise the beam, reducing its space charge and reducing the beam blow-up. Beams can be almost 100% space charge compensated, meaning they see almost no Coulomb force beam blow-up.

Sometimes a heavy 'buffer gas' suck as krypton or xenon is deliberately bled into the vacuum vessel to provide a background gas for the beam to ionise.

Space charge compensation is a complex dynamic process, for pulsed ion beams it can take about 100 µs to build up the compensation particles so the start of a pulsed beam will have a transient change in emittance. For very long beam pulses (> 1 ms) the beam can actually lose its compensating particles by diffusion, a process known as decompensation.

Magnetic fields complicate the spatial and temporal characteristics of space charge compensation further by magnetically confining the compensating particles.

Space charge compensation is insignificant in accelerating gaps because any compensating particles produced are swept out of the gap by the field. For the short time compensating particles remain in the gap they are not effective at compensating the beam because they are moving too fast.

### 3.8 Child-Langmuir Law

There is an absolute limit to the current density that can be extracted from a plasma. There comes a point where the space charge of the beam being extracted actually cancels out the extraction field, making it impossible to extract a higher current density without increasing the extraction field. The current density where this happens can be calculated from the Child-Langmuir equation:

$$j = \frac{\frac{4}{9}\epsilon_0 \sqrt{\frac{2q_i}{m_i}} V^{\frac{3}{2}}}{d^2} , \tag{13}$$

where:

– $j$ is the current density in Am$^{-2}$

- $q_i$ is the ion charge in coulombs
- $d$ is the extraction gap in meters
- $m_i$ is the ion mass in kg
- $V$ is extraction voltage in volts.

The Child-Langmuir equation assumes a uniform electric field and an axisymmetric beam. It does not take into account the voltage drop across the apertures in the electrodes, it does take into account the aspect ratio of the aperture (higher currents can be extracted from a slit compared to a circular aperture of the same area). If more useful units are substituted Eq. (13) becomes:

$$j = \left(\frac{1.72\sqrt{\frac{q}{A}}}{d^2}\right) V^{\frac{3}{2}}, \qquad (14)$$

where:

- $q$ is the ion charge state
- $A$ is the ion mass in atomic mass units
- $d$ is the extraction gap width in cm
- $V$ is the extraction voltage in kV
- $j$ is the current density in mAcm$^{-2}$.

These equations are true when the particle source is running in 'space charge limited' mode, i.e. where the plasma generator has plenty of particles to give, but space charge limits the extracted current. If the plasma cannot give any more particles then the source is in 'plasma limited' or 'source limited' mode and the current vs voltage relationship shown in Eq. (14) no longer follows.

### 3.9 Perveance

The perveance, $P$ of particle source is defined as:

$$P = \frac{I}{V^{\frac{3}{2}}}, \qquad (15)$$

where $I$ is the beam current.

If the particle source is running in 'space charge limited' mode, $P$ should be constant as the extraction voltage is increased. If $P$ starts to decrease then this is an indication that the plasma can no longer supply enough particles to the extractor, the source moves into 'plasma limited' or 'source limited' mode. The word perveance comes from the Latin 'pervenio' meaning to attain. Perveance is also called 'puissance' in some texts, this is actually a better word as it is French for strength or ability, and perveance is a measure of the strength or ability of the source to deliver ions to the extraction system.

### 3.10 Pierce Electrode Geometry

The shape of the electric field in the extraction gap will shape the beam as it is extracted. The Pierce electrode geometry is an attempt to produce an absolutely parallel beam. Named after John Pierce who developed the concept at Bell Labs. The idea is to produce an extraction field that has a zero transverse value at the edge of the beam, thus not having any focusing effect on the beam. The standard Pierce geometry consists of a plasma electrode with an angle of 67.5° to the beam axis (22.5° to the perpendicular) as shown in Fig. 6. The extraction electrode is curved along an equipotential line to the solution of the equation that gives zero traverse field at the beam edge. The Pierce extraction geometry can be applied to circular and slit apertures.

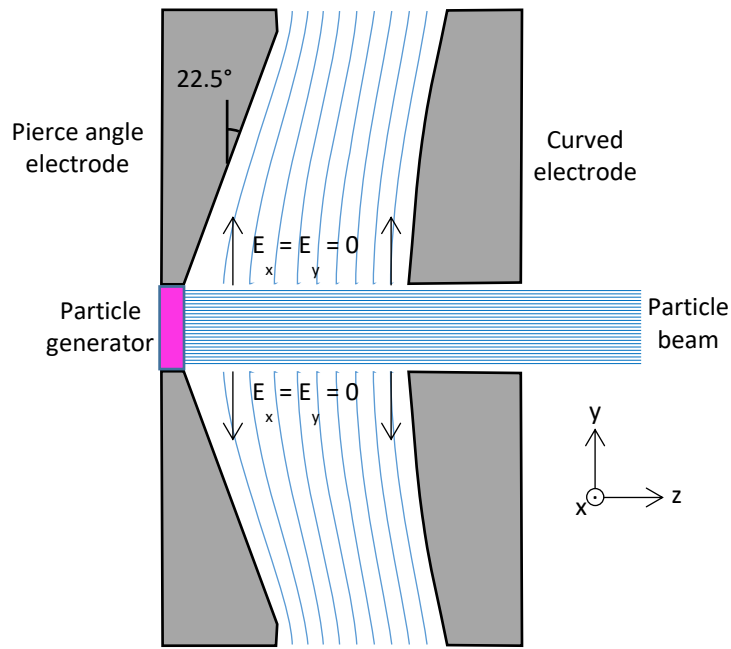

**Fig. 6:** The Pierce extraction electrode geometry.

In reality a completely parallel beam is impossible because of space charge effects.

## 3.11 Suppressor Electrode

In accelerating gaps for positive ions, the electrons will be accelerated in the opposite direction and into the ion source. This is not desirable so an electron suppressor electrode is often added just before the ground electrode. The suppressor electrode is biased slightly more negative than the ground electrode. Any electrons heading into the acceleration gap from the ground electrode side will be reflected back as shown in Fig. 7.

In negative ion sources, protons will be accelerated back into the source instead of electrons so the suppressor is biased with a positive voltage. Back streaming particles can damage the source by sputtering.

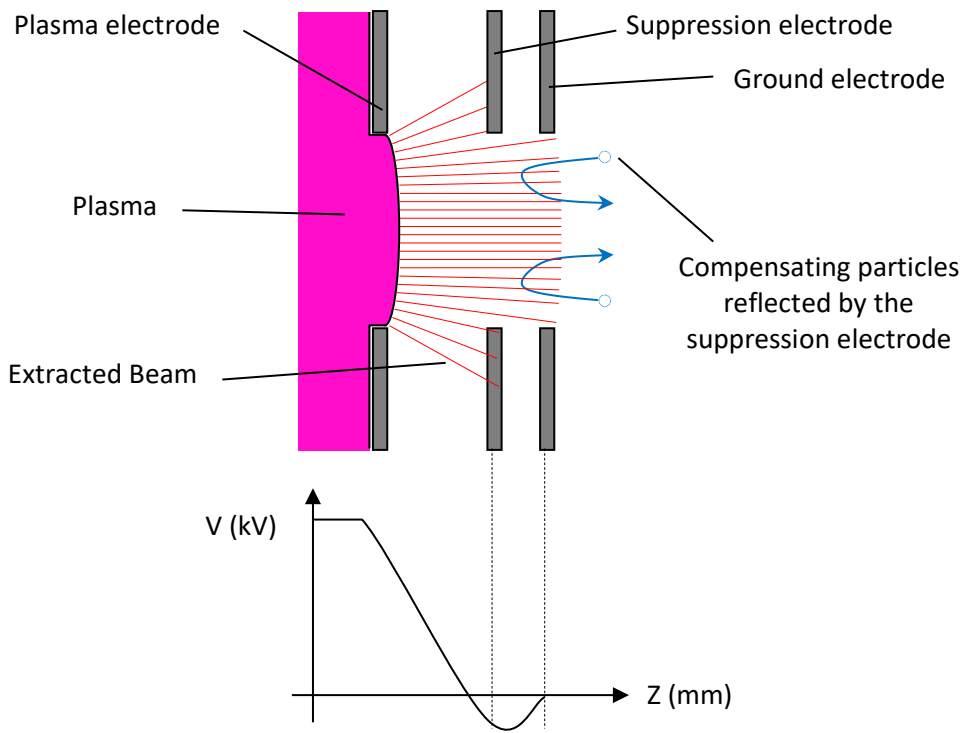

**Fig. 7:** The use of a suppressor electrode to prevent back streaming particles of the opposite charge entering the ion source.

### 3.12 Negative Ion Extraction

Extracting electrons with negative ions is obviously unavoidable because they both have the same charge. Dealing with the co-extracted electrons is one of the main challenges for high current negative ion source extraction systems. The ion source designer must first try to minimise the amount of co-extracted electrons, then find a way to separate and dump the unwanted electrons from the negative ion beam. In some cases the electron current can be 1000 times greater than the negative ion current itself. For high duty factor sources a cooled electron dump is required.

### 3.13 Low Energy Beam Transport

The extraction system starts at the particle generator and ends when the beam reaches the energy required by the next stage of the accelerator. The beamline between the extraction system and the next stage of the accelerator is called the Low Energy Beam Transport (LEBT). Sometimes the exact point where the LEBT starts can be a matter of local terminology or history. A post-extraction-acceleration stage might come between the extraction system and the LEBT.

The whole ion source usually sits on a high voltage platform. The beam is accelerated to ground (this is why the last electrode of the extraction system in Fig. 7 is labelled as the ground electrode), the beam then enters the LEBT.

LEBTs can be magnetic or electrostatic or a combination of both. It is common to use between 1 and 4 focussing elements. These can be electrostatic Einzel lenses or electromagnetic solenoids and quadrupoles. In negative ion sources that use a large amount of caesium vapour electromagnetic LEBT can be used to reduce sparking problems. The LEBT beam optic design must transport and match the beam into the acceptance of the next phase of the accelerator.

To give representative beam current measurements, ion source test stands should include some of the LEBT. Beam halo and emittance effects mean that the current measured directly after extraction is not a true measure of the beam current that can be transported to the next stage of the accelerator. Often there is significant collimation of the beam on the way through the LEBT and a large proportion of the beam current can be lost. Higher pressures in the LEBT can lead to beam current loss and emittance growth. The degree of space charge compensation is affected by the vacuum pressure. For negative beams higher pressures lead to increased beam loss through stripping.

# 4 Electron Guns

## 4.1 Brief Historical Introduction

The first electron beams were produced by William Crookes and his assistant Charles Gimingham in London in the 1870's using an improved version of Hermann Sprengel's mercury displacement vacuum pump and a glow discharge tube developed by Heinrich Geißler. Gimingham's improved pump allowed Crookes to evacuate the glass Geißler tubes to pressures as low as $10^{-5}$ mBar. At these pressures the glow discharge went out, leaving a shadow of the anode in the scintillation on the glass wall behind the anode. Crookes interpreted this as radiation emitted from the cathode, calling the effect cathode rays. Geißler's tubes and Sprengel's pumps were also used by Julius Plücker, Johann Hittorf, Eugen Goldstein, Heinrich Hertz, Philipp Lenard, Ivan Puluj, Kristian Birkeland and others to discover the properties of cathode rays, culminating in J.J. Thomson's 1897 identification of cathode rays as negatively charged particles, which were later named electrons. Interest in developing electron devices exploded and created the modern world we live in.

Many of the early vacuum tube electronic devices have now been superseded by solid-state silicon versions, but electron guns are still used for high power devices, x-ray machines and electron microscopes. The thermionic cathodes used in these modern devices wear out and several manufacturers produce replacements. These commercially available cathodes are used in thermionic electron guns for particle accelerators.

Electron guns for particle accelerators generally employ either thermionic or photo emission processes to produce the electrons. The electrons are then accelerated by either DC or RF electric fields. The applied DC electric field might be pulsed, but it is steady state for the duration of the pulse. The following sections give examples of combinations of these schemes used in particle accelerators, but it is not exhaustive.

## 4.2 Thermionic Electron Guns

### 4.2.1 Planar Triodes

The basic design of a planar triode thermionic emission gun is shown in Fig. 8. The cathode is made of a low work function material, commonly a sinter of tungsten and barium oxide. An electrical heating element is used to heat the sinter which is designed to slowly release barium oxide thus maintaining a low work function surface at a high temperature. This release process puts a limit on the lifetime of this type of 'dispenser cathode' from days to decades depending on operating conditions. The cathode is positioned in a Pierce extraction gap with an accelerating voltage of 50 - 100 kV applied with the anode at ground potential.

The beam can be pulsed on and off very quickly (ns regime) by applying a voltage to a grid of wires directly in front of the electron emitting surface. This is exactly the same principle used in a vacuum tube triode amplifiers.

Very short 'pancake' pulses can be produced that allow the Child Langmuir limit to be exceeded.

Several manufacturers produce planar cartridge assemblies consisting of the dispenser cathode, heater and grid that can quickly and easily be replaced. CPI Eimac manufacture the YU171 which is a planar triode package used by Diamond Light Source, PSI Swiss Light Source, and BNL among others.

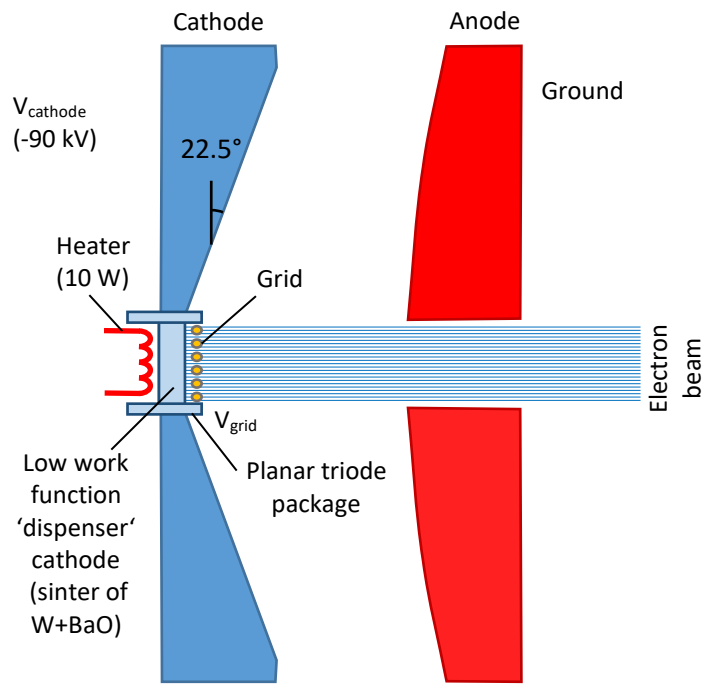

**Fig. 8:** Schematic of a gridded thermionic emission gun with Pierce geometry used in particle accelerators.

### *4.2.2  Single Gap Accelerators*

Relativistic beams are stiffer and less susceptible to beam space charge emittance blow up. For very high pulsed current applications such as Free Electron Lasers (FELs) it is therefore highly desirable to make the beam relativistic as soon as possible. Due to their low mass (511 keV), electron beams become relativistic very quickly. This acceleration can be achieved in a single accelerating gap. Accelerating voltages in the order of 500 kV are achievable but require careful system design.

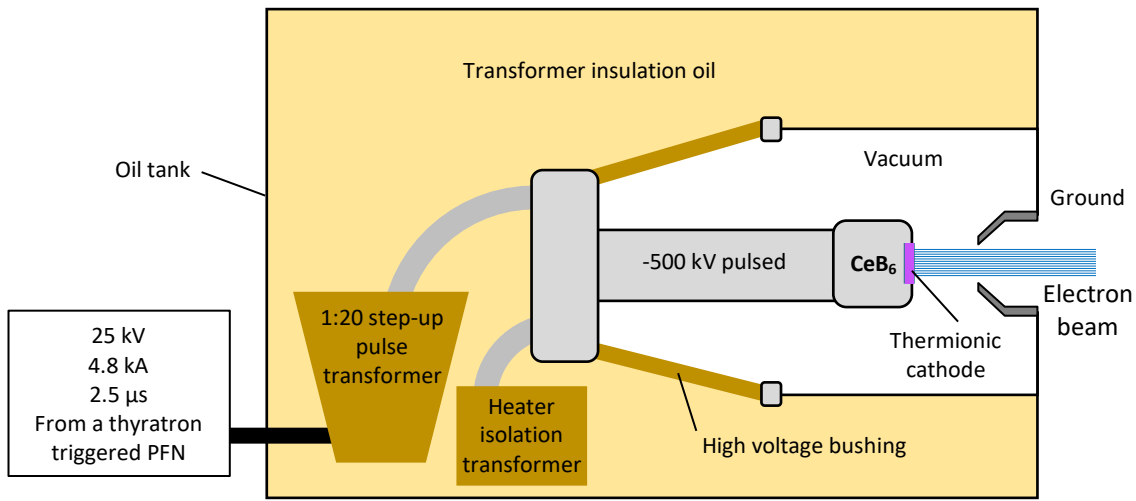

**Fig. 9:** Schematic of the SCALA oil insulated pulsed DC thermionic electron gun.

Figure 9 shows a schematic of the approach used for the SACLA accelerator at RIKEN. An oil insulated pulse transformer is used to step up a -25 kV pulse to -500 kV applied to the cathode. The cathode is made of low work function cerium hexaboride (used in electron microscopes) heated to 1500°C using a heater powered from an isolation transformer. Cerium hexaboride can and can provide long lifetimes of over 1 year.

## 4.3 Photoemission Electron Guns

### 4.3.1 Introduction

Designs for FELs and future linear colliders demand ever higher beam current and brightness, so for many years researchers have been exploring the use of photoemission electron guns. Photo emission materials offer high current densities. They work by shining a high power UV laser onto the photocathode material. Using a laser beam allows very precise control over pulse length and time structure which is essential for accelerators.

Common photocathode materials are $Cs_2Te$ and GaAs. The cathode material often needs special preparation and handling.

### 4.3.2 Single Gap DC Photoemission Guns

A typical setup of a single gap DC photoemission gun is shown in Fig. 10. The photocathode material is mounted in an electrode held at a minus a few hundred kV. The high voltage bushing is a critical part of the design. Some designs have an inverted, re-entrant bushing where the insulator goes into the vacuum vessel instead of coming out of it. For voltages above 200 kV with an air/vacuum bushing, a stacked insulator design is preferable because it allows voltage grading (using external resistive components) and field grading (using guard rings) along the length of the bushing and splits the insulator into shorter sections that are easier to manufacture and handle.

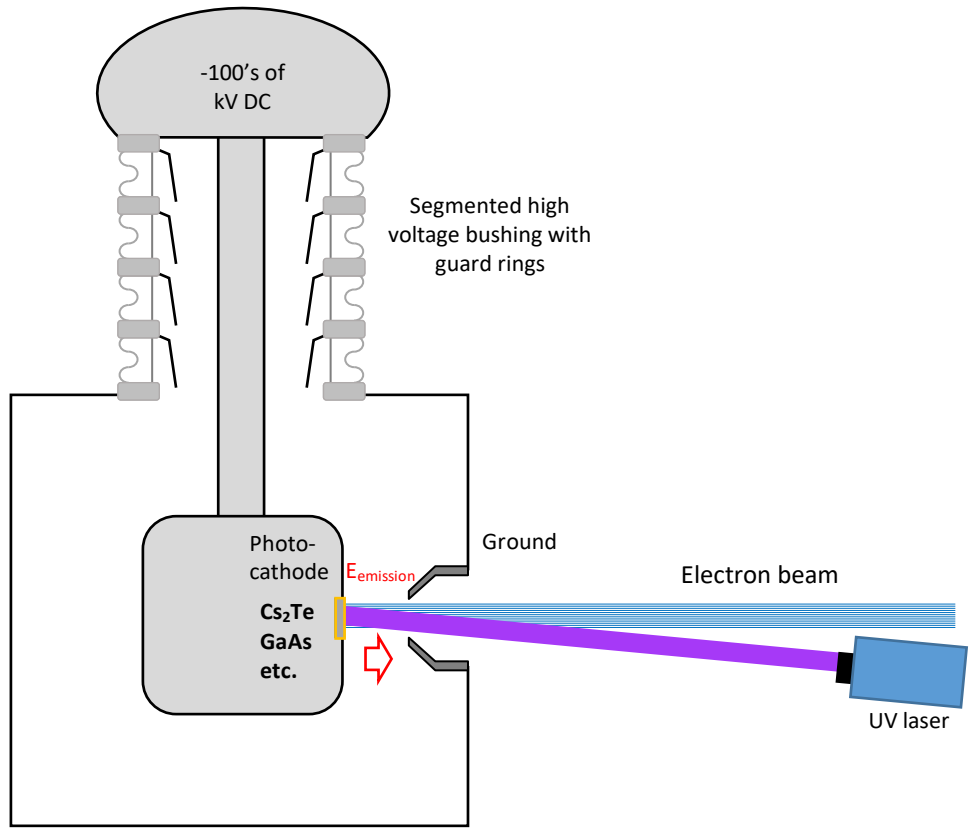

**Fig. 10:** Schematic of a single gap photo emission gun used in particle accelerators.

A high power UV laser is aimed at the photocathode material. The duty cycle of the laser defines the time structure of the beam. To produce polarised electron beams special strained photocathodes and circularly polarised lasers are used.

The type of photocathode material used depends on the current and duty cycle required. Thermal considerations play a key part of the design. Also, the method of changing the photocathode must be carefully considered, photocathodes can have short lifetimes and some must always be kept under vacuum, if the photocathode takes too long to change is has consequences for the availability of the overall machine.

*4.3.3    RF Photoemission Guns*

RF photoemission guns were developed for FELs in the 1980s but are now being developed for other high current applications such as compact gamma-ray or X-ray sources based on inverse Compton scattering, or as direct electron sources for ultrafast time-resolved pump-probe electron diffraction.

The basic design of an RF photoemission gun (Fig. 11) is similar to the DC gun in the previous section, in that it employs a photocathode and a UV laser. The difference is that the accelerating field on the surface of the photocathode ($E_{emission}$) is produced by an RF cavity. The cavity has n + ½ cells (where n is an integer) so that the peak longitudinal field that is normally in the middle of a cell is positioned on the surface of the photocathode in the ½ cell. The RF cavity can be either normally or superconducting and operates at GHz frequencies. The cavity and photocathode is surrounded by a solenoid to confine the beam. Much higher fields (> 10 MVm$^{-1}$) can be achieved in an RF gun.

It is much easier to cool the photocathode in an RF gun because it is at ground potential.

The laser produces very short pulses that allow 'pancake' beam pulses to be perfectly positioned on the rising edge of the peak of the applied RF waveform as shown in Fig. 11.

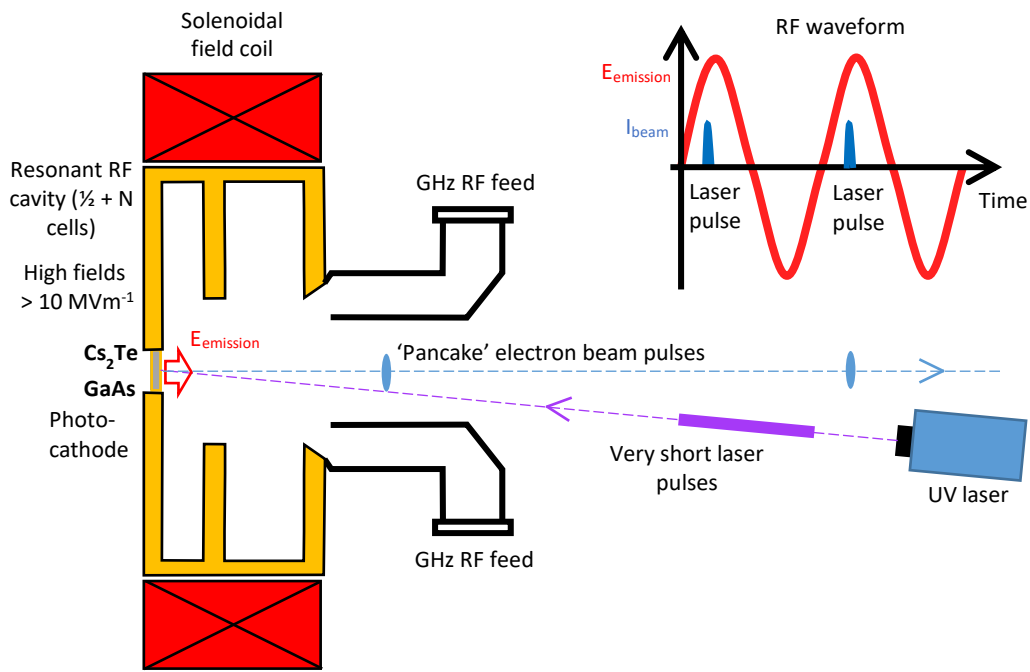

**Fig. 11:** Schematic of an RF photoemission gun used in particle accelerators.

# 5 Positive Ion Sources

## 5.1 Introduction

Researchers have been experimenting with beams of positive ions (or 'canal rays' as they called them) since 1886 when Eugen Goldstein discovered that they were emitted from holes in the cathode. The positive ions in the plasma were being accelerated by the cathode sheath potential, and emerging as beams through the holes in the cathode. The problem with 'canal ray' sources is that the beam energy is defined by the discharge voltage and the beam has a huge energy spread (as large as the discharge voltage).

## 5.2 Hot Cathode Sources

### *5.2.1 Basic Filament Source*

The first practically useful positive ion source was developed by Arthur Dempster at the University of Chicago in 1916 [2]. It was the first source to introduce an extraction electrode and have the ions emitted from a hole in the anode which radically reduces the energy spread of the ions produced. The basic design of a hot cathode filament ion source is shown in Fig. 12.

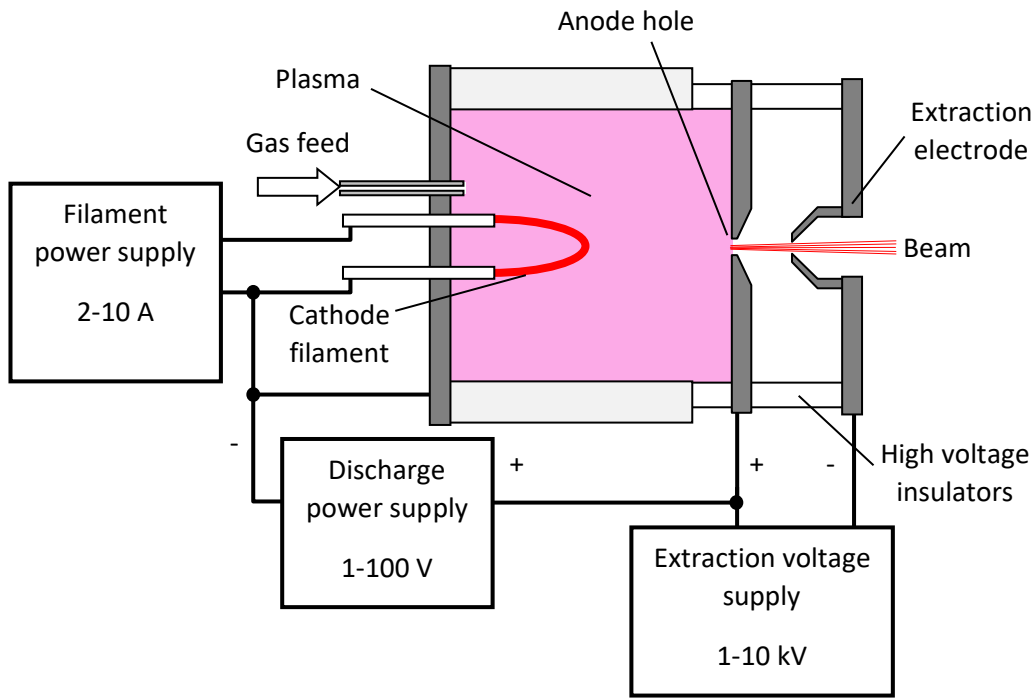

**Fig. 12:** Schematic of a hot cathode filament source.

A current is passed through the cathode filament to heat it to the point where it thermionically emits electrons. Coil filaments can be used but the hairpin or 'U' shape are more common. The thermionically emitted electrons are accelerated across the plasma sheath field that exists at the cathode, created by the voltage applied by the discharge power supply. The anode has a small hole in it, opposite which there is an extraction electrode. A negative high voltage is applied to the extraction electrode. The positive ions produced near the anode hole are extracted from the source. The beam current can be increased by increasing the discharge current and/or extraction voltage. Beam currents of about 1 mA can be produced before dedicated cooling is required.

Filament sources are cheap and easy to manufacture. They can be used to produce positive ions with low charge state of a wide range of elements.

### *5.2.2 High Current Filament Sources*

For higher beam currents a multicusp magnetic field is often added around the chamber to confine the plasma. Also, a larger diameter filament wire is used which could require several hundred amps of DC to heat it up so that it thermionically emits electrons. The temperature of the filament should not be too high otherwise the electron emission will become space charge limited. This causes the main discharge voltage to increase and become very noisy.

The shape and position of the hot cathode filament is important. Wide hairpins ('U'shaped) have been shown to produce more stable plasmas over a large range of operating conditions. The position of the filament relative to the multicusp field is important because the high current required to heat the filament loop creates a magnetic field itself. The filament should therefore be positioned so that the filament produced magnetic field is in the same direction as the local cusp field. Sometimes more than one filament is used.

Discharge currents (usually pulsed) in the multiple hundred amp range are required to produce very high beam currents. As discharge currents increase more care must be paid to the cooling system design.

### *5.2.3 Plasmatrons*

The plasmatron was developed by the prolific aristocratic German inventor Manfred von Ardenne in the late 1940s. It is a development of the basic filament source. To increase the beam current, a conical shaped intermediate electrode is positioned between a heated filament cathode and an anode as shown in Fig. 13. The purpose of the conical intermediate electrode is to 'funnel' the plasma down to a higher density region near the anode extraction hole. A plasma double sheath forms on the conical intermediate electrode. The higher plasma density near the extraction region allows more ions to be extracted.

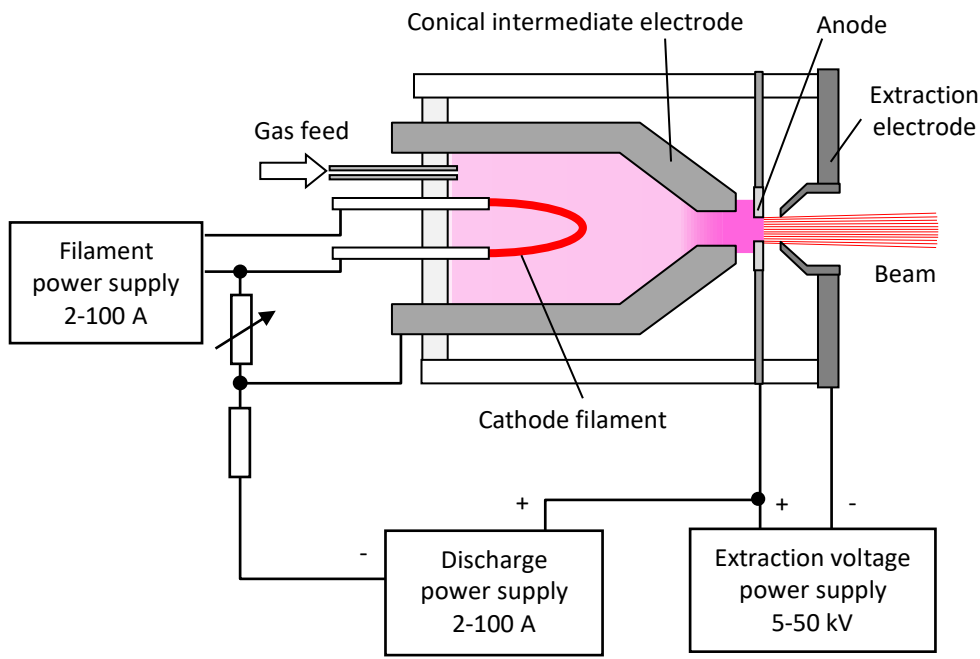

**Fig. 13:** A schematic of a plasmatron source.

*5.2.4 Duoplasmatron*

Von Ardenne continued to develop the plasmatron and in 1956 he invented the Duoplasmatron, shown in Fig. 14. It is effectively the same as a plasmatron but the conical intermediate electrode is made of soft iron. The source is positioned inside a solenoid. The conical soft iron intermediate electrode squeezes the axial magnetic field lines and concentrates them just in front of the anode. The squeezing of the magnetic field lines and the funnelling effect of the cone create a very high plasma density just in front of the extraction hole. This greatly increases the ion density and allows very high ion currents of up to 500 mA to be extracted. The ions streaming through the anode hole are too dense to allow the extraction of beams with a uniform distribution and low emittance so the plasma is allowed to expand into an expansion cup before being extracted. To prevent back streaming electrons a suppressor electrode is used. The name duoplasmatron comes from the two very different plasma densities that exist in the source.

The duoplasmatron a common type of high current positive ion sources because it is cheap and easy to maintain. Lifetimes are limited to a few weeks at high currents because of filament erosion, but filaments can be easily replaced. At lower currents lifetimes can be as long as several months.

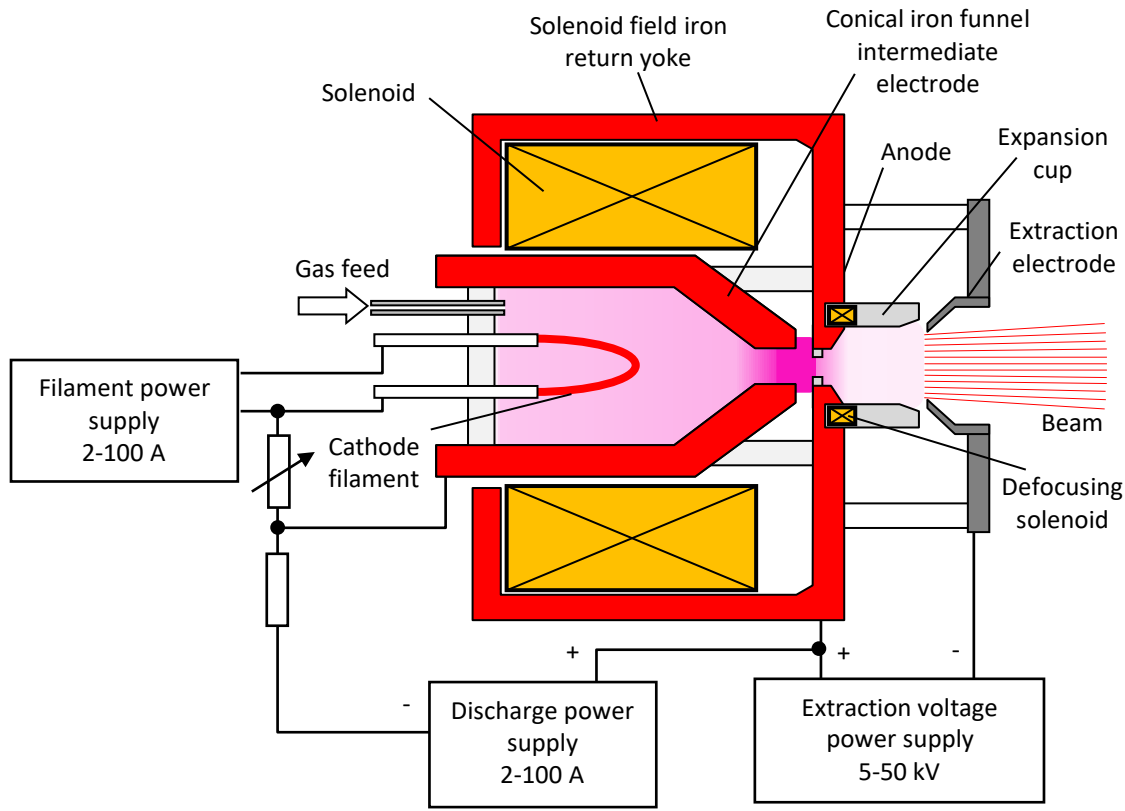

**Fig. 14:** A schematic of a duoplasmatron source.

## 5.3 Cold Cathode Sources - Magnetron and Penning

*5.3.1 Introduction*

The term 'cold cathode' refers to the fact that the cathode is not independently heated (like the filaments in the previous section), however the name can be misleading because the cathode can still operate at an elevated temperature due to heating by the discharge itself. In these types of sources the plasma discharge is in direct contact with the anode and cathode, so sputtering processes will eventually erode the electrode surfaces. This puts a fundamental limit on the lifetime of these sources. However, they are relatively simple to manufacture and the consumable parts can easily be replaced. They can be used to produce positive beams with low charge state of a wide range of elements.

*5.3.2   Fundamental Geometry*

The fundamental topologies of the magnetron and Penning electrode geometries are shown in Fig. 15. The magnetron geometry consists of a cathode surrounded by an anode. The Penning geometry consists of a tubular anode with cathodes facing each other at either end of the anode. In both geometries the magnetic field is in line with the axis and the beam is extracted through a hole in the anode (not shown).

Figure 15 shows the fundamental electrode topologies, but is likely their actual shapes will be different in reality: the anode might be square or rectangular instead of round; the anode might be very short, like a window frame; the cathode in the magnetron might be as thin as a wire; the electrodes might be comprised of several separate components held at anode and cathode voltage.

Occasionally the anode and cathode are polarities are swapped in a magnetron, the geometry is then called an inverse-magnetron.

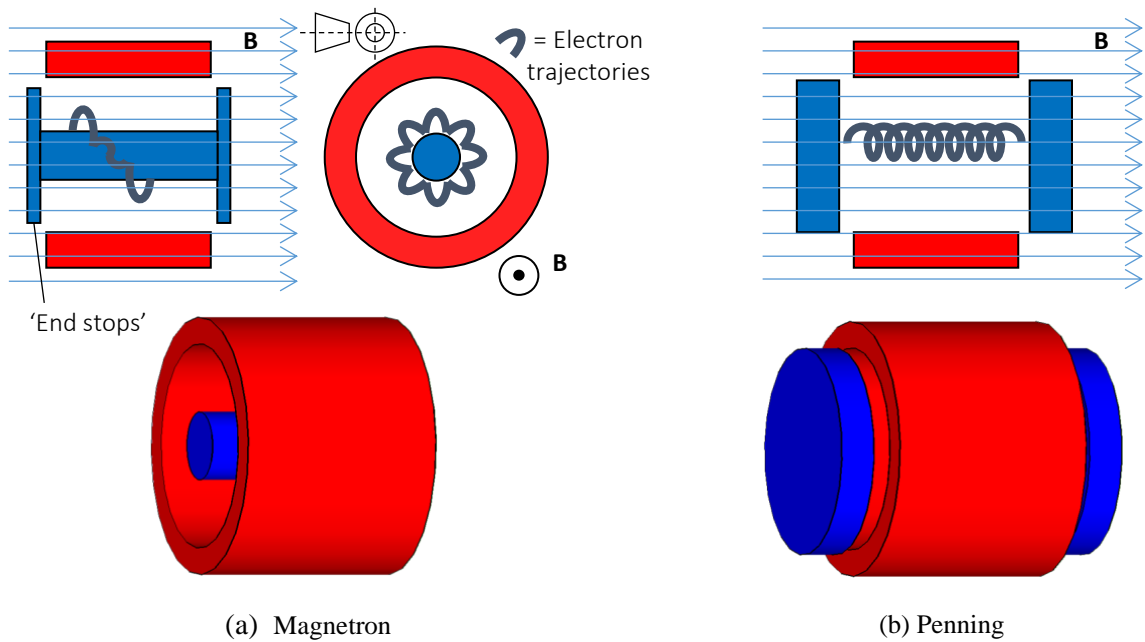

(a) Magnetron  (b) Penning

**Fig. 15:** The fundamental topologies of the magnetron and Penning geometries.

*5.3.3   Magnetron*

The magnetron electrode topology was first invented, not as an ion source, but as an electron tube by Albert Hull working at the General Electric Research Laboratory, Schenectady, New York in 1916 [3]. He was trying to use magnetism to find alternatives to the patented electrostatic control of valve amplifiers. The tubes he invented had various names, such as comet valves, boomerang valves and ballistic valves because of the electron trajectories in the tube (see Fig. 15(a)), however the name magnetron valve eventually stuck [4]. By adding 'end-stops' (see Fig. 15(a)) to the central cathode 'spool' the electrons can also be confined at the ends of the cathode The ion beam is extracted through a hole in the anode.

The magnetron geometry has been developed by many researchers over the decades. The geometry is used in the Freeman source for industrial ion implantation. The magnetron geometry is employed in caesiated negative ion sources (see Section 6.5.1) where it is also sometimes referred to as a 'planotron'.

*5.3.4    Penning*

The Penning geometry was first reported as an ion source by Louis Maxwell working at the Franklin Institute in Philadelphia in 1930 [5]. However, the Penning electrode geometry gets its name from Frans Penning, a researcher working at Philips Physics Laboratory in Eindhoven, The Netherlands. In 1937 he developed the Penning Ionization Gauge or Philips Ionization Gauge (PIG) [6]. By measuring the discharge current in the electrode geometry, he was able to accurately and reliably measure very low pressures in gases, a technique that is still widely used today. Confusingly, Penning ion sources are also referred to as PIGs and electrons bouncing between two cathodes along magnetic field lines is called PIGging or reflexing, hence the alternative name: a reflex geometry or reflex discharge. The ion beam is extracted through a hole in the anode.

The Penning geometry has been developed by many researchers over the decades. The geometry is employed in Calutron, Bernas and Nielson sources used in isotope separation and surface implantation. Sometimes one of the two cathodes can be independently heated. The Penning geometry is used in negative ion sources (see Section 6.5.2).

## 5.4    Vacuum Arc Ion Sources

Instead of creating a plasma by ionising a gas fed into the plasma chamber, vacuum arc ion sources vaporise and ionise the cathode material to produce a plasma. Vacuum arc ion sources are ideal for creating low charge state beams of metals and other normally solid elements. They are fundamentally lifetime limited because the cathode is consumed during operation.

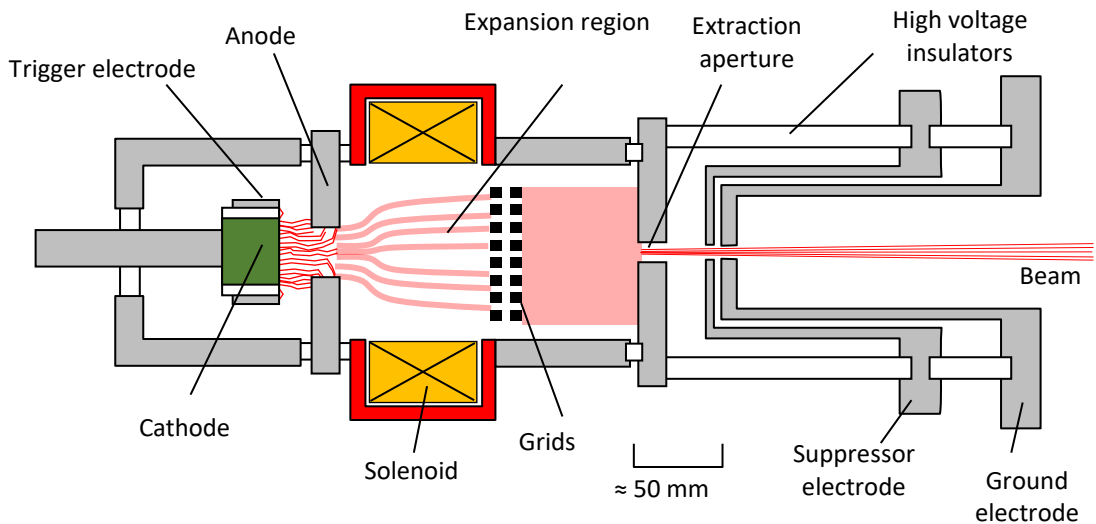

**Fig. 16:** A schematic of a Vacuum Arc ion source.

The cathode in a Vacuum Arc source is surrounded by a trigger electrode as shown in Fig. 16. When a high voltage is applied to the trigger electrode it initiates arc discharges between the cathode and anode. Many hundreds of individual transient arc discharges sputter and vaporise the cathode material. Plasma from these arc discharge 'beamlets' passes through a hole in the anode into an expansion region in a solenoidal field as shown in Fig. 16. The expanding plasma then passes through a positively biased grid which removes the electrons, the space charge forces of the remaining positive ions causes the 'beamlets' to rapidly spread out creating a homogenous distribution that can be extracted as a relatively noise free beam.

Vacuum Arc sources are often called MEVVA (MEtal Vapour Vacuum Arc) sources. The Lawrence Berkley Lab and GSI MEVVA's can produces 15 mA of $U^{4+}$ ions.

## 5.5 Laser Ion Sources

Like vacuum arc ion sources, laser ion sources are good for creating beams of metals and other normally solid elements. A high power pulsed laser (1-100 J per pulse) is aimed through a window at a target made of the material to be ionized. The laser power density vaporizes and ionizes the target material creating a plasma plume that drifts into an expansion region as shown in Fig. 17. The beam is then extracted through a hole. To try to minimize pulse to pulse variability a fresh area of target is used for each shot, this is achieved by having a rotating target or a ribbon of target material.

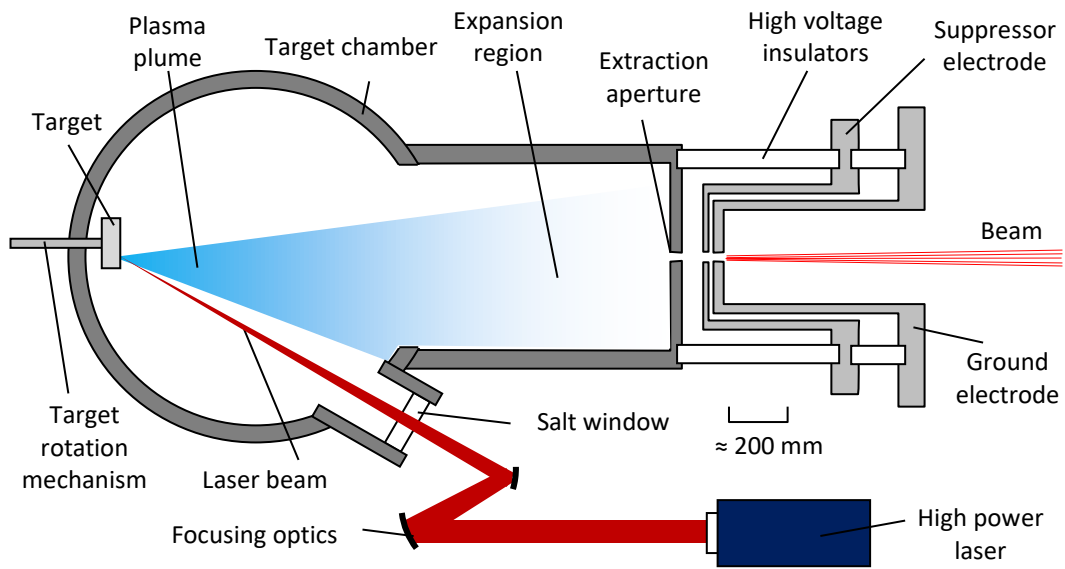

**Fig. 17:** A schematic of a Laser Ion Source.

The TWAC accelerator at ITEP Moscow uses a laser ion source to produce 7 mA, 10 µs pulses of $C_4^+$ ions. Masahiro Okamura, BNL and RIKEN has demonstrated direct plasma injection into an RFQ from a laser ion source.

## 5.6 RF Ion Sources

### 5.6.1 Introduction

The first RF ion source was first developed by Peter Thonneman [7] at the University of Sydney in 1946 for his Master's degree. Since then it has been developed by many researchers, and the frequencies employed have expanded from MHz to 10's of GHz.

In RF sources the high frequency power must be coupled to the plasma, this is achieved by transferring energy to the electrons which can be done either inductively or capacitively. For time varying fields both inductive and capacitive coupling mechanisms always exist. Capacitively coupled sources employ the magnetron geometry to generate the electric field that accelerates the electrons by applying an RF voltage to the cathode electrode. Inductively coupled sources rely on the time varying electric field induced by the time varying magnetic field produced by the current in a coil to accelerate the electrons. At microwave wavelengths it becomes feasible to use waveguide to couple into the plasma chamber.

*5.6.2    Inductively Coupled Plasma Sources*

Inductively Coupled Plasma (ICP) sources work in the MHz frequency range. The power is transferred to the plasma via an alternating current in a coil of a few turns. The RF coil is often referred to as an antenna, but this is not technically correct because it is not transmitting an electromagnetic wave. In the MHz regime the wavelength is several orders of magnitude longer than the coil dimensions, the electrons predominantly interact with the time varying field inside the coil (which is definitely the 'near field'). The accelerated electrons impact the heavy atoms, molecules and ions, thus driving the plasma.

The coil can be positioned inside or outside the plasma as shown in Fig. 18. A coil internal to the plasma chamber provides the highest coupling to the plasma because the entire bore of the coil is occupied with plasma as are volumes outside the coil where capacitive coupling dominates. The internal coil must be coated in a layer of insulating ceramic which is subject to sputtering erosion by the plasma, the source fails when the plasma finally eats through this layer. A coil external to the plasma chamber requires the plasma chamber to be made out of insulating material, usually an alumina ceramic. With the coil behind an insulator and no electrodes to erode away, external coil RF ion sources can have very long lifetimes in excess of 1 year. The plasma chamber is often put in a solenoidal or multicusp magnetic field to confine the plasma.

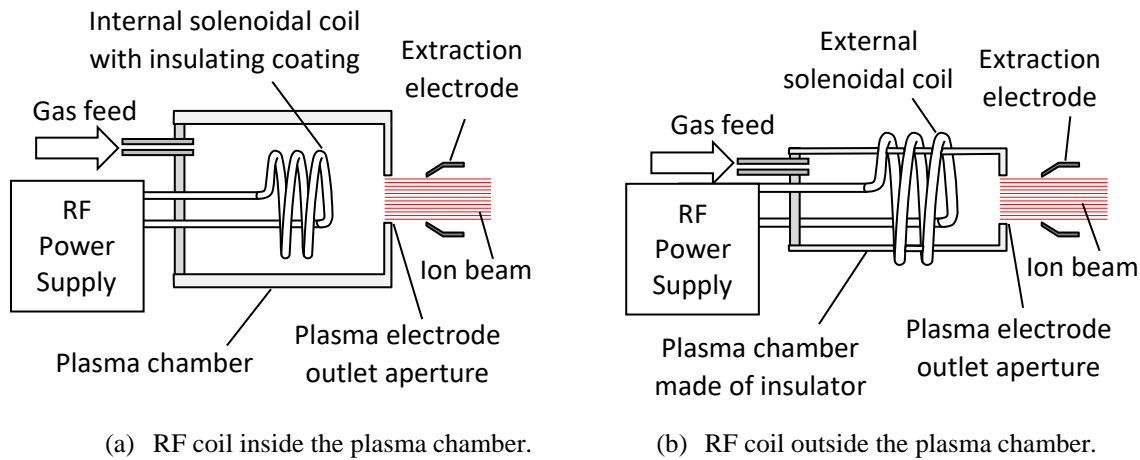

(a)  RF coil inside the plasma chamber.     (b)  RF coil outside the plasma chamber.

**Fig. 18:** The two options for a coil in an inductively coupled RF Source.

ICP RF sources have been developed by numerous researchers over the decades with various coil configurations including flat planar spirals both inside and outside the plasma chamber. ICP RF plasmas are also used in negative ion sources.

*5.6.3    Microwave Discharge Ion Sources*

Microwave discharge ion sources operate in the GHz frequency range to generate the plasma. The microwave energy is coupled to the discharge via an antenna or waveguide. With an antenna behind an insulator or a waveguide there are no electrodes to erode away, so microwave ion sources can have very long lifetimes in excess of 1 year. The plasma chamber has similar dimensions to the wavelengths of the microwaves and is surrounded by DC solenoids that produce an axial magnetic field.

Microwave discharge ion sources can produce very high currents of singly charged ions with low emittance. Microwave discharge ion sources were first developed by Noriyuki Sakudo's team at Hitachi [8] and Junzo Ishikawa's team at Kyoto University [9] in the late 1970s and early 1980s. The basic design of all modern microwave discharge ion sources are based on the proton source developed by Terence Taylor and Jozef Mouris at Chalk River National Laboratory in the early 1990s [10].

Figure 19 shows a schematic of a microwave ion source. The key aspects of the design are a small compact plasma chamber with two solenoids at the front and back that confine the plasma. A stepped matching section is used to allow smooth transition between the wave guide and plasma impedances. An extraction system with a suppressor electrode is employed to limit the back streaming electrons.

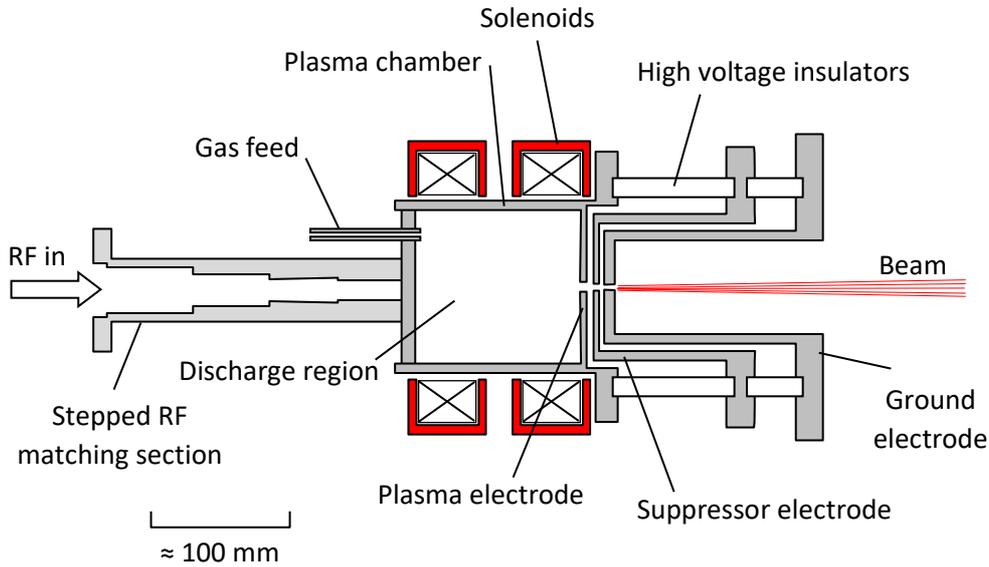

**Fig. 19:** Schematic of a microwave discharge ion source.

The design was further improved by Joe Sherman and team at LANL in the mid 90s [11]. During the second half of the 1990s Rafael Gobin and team at CEA Saclay, developed the SILHI (Source d´Ions Légères a Haute Intensités) source [12,13]. They obtained greater brightness and even higher reliability. Proton beam currents of 140 mA with 0.2 π mm mrad normalised emittance were demonstrated. Lifetimes of around one year were demonstrated. The SILHI source design is being used for the European Spallation Source.

### 5.6.4   *Electron Cyclotron Resonance (ECR) Sources*

When the magnetic field and frequency are just right in a microwave source, the electrons can be cyclotron accelerated. ECR ion sources (Fig. 20) were first developed in the 1970s by Richard Geller and his team at CEA. ECR ion sources can produce multiply charged positive ions. ECR sources work by step-by-step ionisation: the accelerated electrons progressively remove the outer orbital electrons from ions by impact ionisation. The comparatively slow moving positive ions are confined by the magnetic fields to be ionised multiple times. This creates high charge state positive ions and allows the production of high charge state beams of heavy elements such as lead or uranium. ECR sources produce a distribution of charge states, so m/q separation in a dipole downstream is required to select single charge state beams.

The ECR ion source is based on electron heating at the electron cyclotron frequency ($f_{ECR}$) in a magnetic field, given by:

$$\omega_{ECR} = 2\pi f_{ECR} = \frac{eB}{m} \ . \tag{16}$$

For a 2.45 GHz frequency the electron ECR field is 875 G. For electrons in a magnetic field in the range 0.05–1 T, a frequency range of 1.4 GHz to 28 GHz is used. The availability of commercial magnetrons and klystrons results in most sources working at 2.45, 10, 14.5, 18, 28 and 37.5 GHz. 2.45 GHz is often used because of microwave ovens and cheap reliable magnetron devices are readily available. Also, the waveguides are of manageable size (35 mm x 73 mm). Lower frequency yields lower emittance beams and requires lower magnetic fields.

The two solenoids produce a 'minimum B' magnetic trap that confines the positive ions. A hexapole field around the plasma chamber also increases confinement. The fields are generated either by permanent magnets or coil windings. The combination of hexapole and solenoidal fields is also important for plasma stability because they create a 'closed ECR surface' where the ionisation takes place. The closed ECR surface is positioned in the middle of the plasma chamber as shown in Fig. 20.

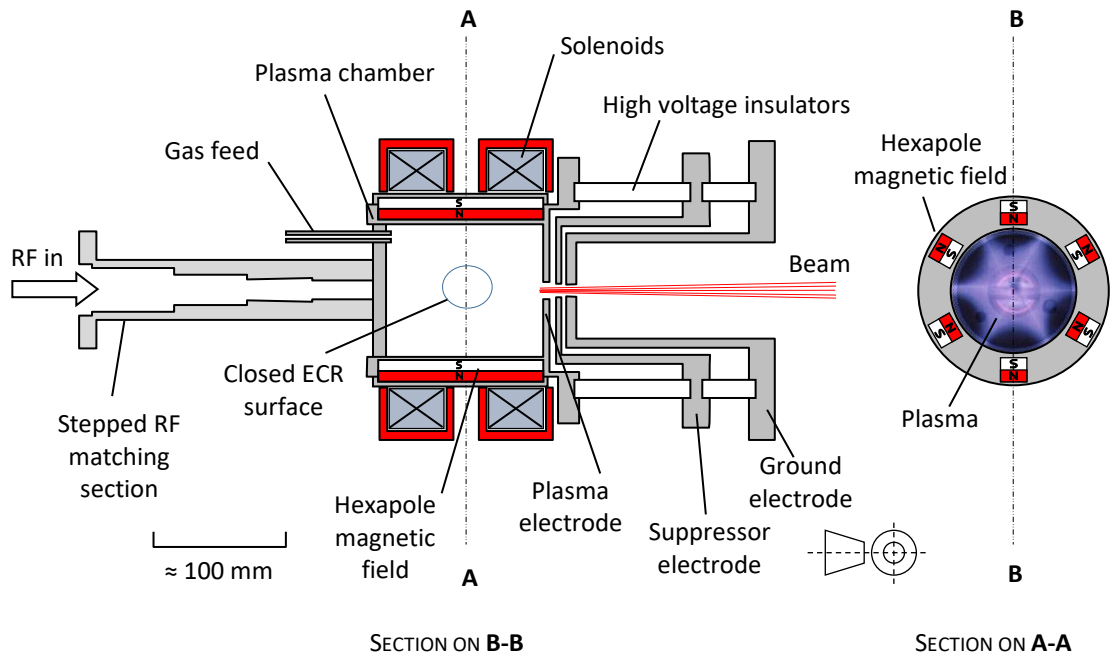

**Fig. 20:** Schematic of an ECR ion source.

ECR sources can be run in continuous or pulsed mode. Pulsed heavy ion ECR sources (such as the GANIL/CERN $Pb^{27+}$ ECR source) run in pulsed 'afterglow' mode where the beam is extracted after the RF has been turned off. Higher current short pulses can be achieved in this way.

To reach the highest currents and charge states, denser plasmas are required. This is achieved with stronger solenoidal fields, stronger hexapole fields and higher RF frequencies. Superconducting solenoids and hexapoles are used in high performance ECR sources. A good example is the 28 GHz superconducting VENUS ECR at Lawrence Berkeley Laboratory which can produce 200 eµA of $U^{34+}$ ions or 4.9 eµA of $U^{47+}$ ions.

ECR sources are used as 'charge breeders' in experiments with radioactive beams. Radioactive elements produced in the target ion source are then injected into an ECR where they are ionised to higher charge states.

### 5.6.5  *Gas Dynamic ECR sources*

Gas Dynamic or Quasi Gas Dynamic ECR sources use pulsed gas and high power, high frequency RF to create a transient plasma discharge capable of producing very high current beams. Along with the duoplasmatron, they are the only ion sources capable of producing beam currents in the multiple 100 mA range. Gas Dynamic ECR sources are a relatively new technology, they have been developed by the Russian Academy of Sciences Institute of Applied Physics, and the University of Jyväskylä, Finland since the 2010s [14].

A schematic of a Gas Dynamic ECR is shown in Fig. 21. In most ion sources, gas is fed into the plasma chamber and then pumped away through the extraction aperture. In a gas dynamic source there are multiple pumps in the plasma chamber to allow a very high pressure transient gas pulse to pass through the source. The gas and RF are injected on beam axis.

Plasma lifetime is determined by the time of the gas dynamic flux of plasma through the solenoidal trap so very high RF frequencies (35-70 GHz) and high fields in the solenoids are used to increase plasma confinement.

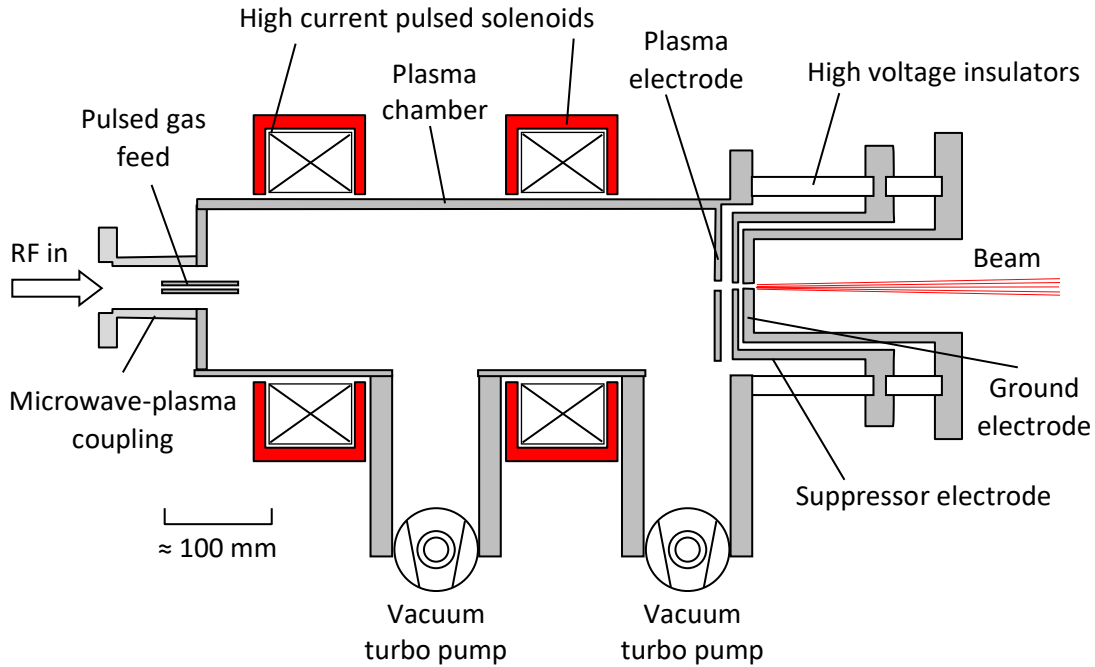

**Fig. 21:** Schematic of a gas dynamic ECR ion source.

## 5.7 Electron Beam Ion Sources

### 5.7.1 Introduction

Electron Beam Ion Sources (EBIS) use a high current density electron beam to ionise the particles. The EBIS was first developed in the late 1960s by E.D. Donets and his team at JINR, Dubna. EBIS are complex, expensive and can only produce relatively short pulse lengths. However, they are capable of reliably producing beams of very high charge state positive ions. Heavy elements can be completely stripped of their electrons leaving bare nuclei.

### 5.7.2 Basic Operation

Figure 22 shows a schematic of an EBIS. A high current electron gun produces a 1-20 keV electron beam that is compressed to a current density in the order of 1000 $Acm^{-2}$. The electron beam passes though through a set of drift tubes in a 1-5 T solenoidal field. The strong solenoidal field radially confines the electron beam. Damping components on the drift tubes help maintain the beam stability.

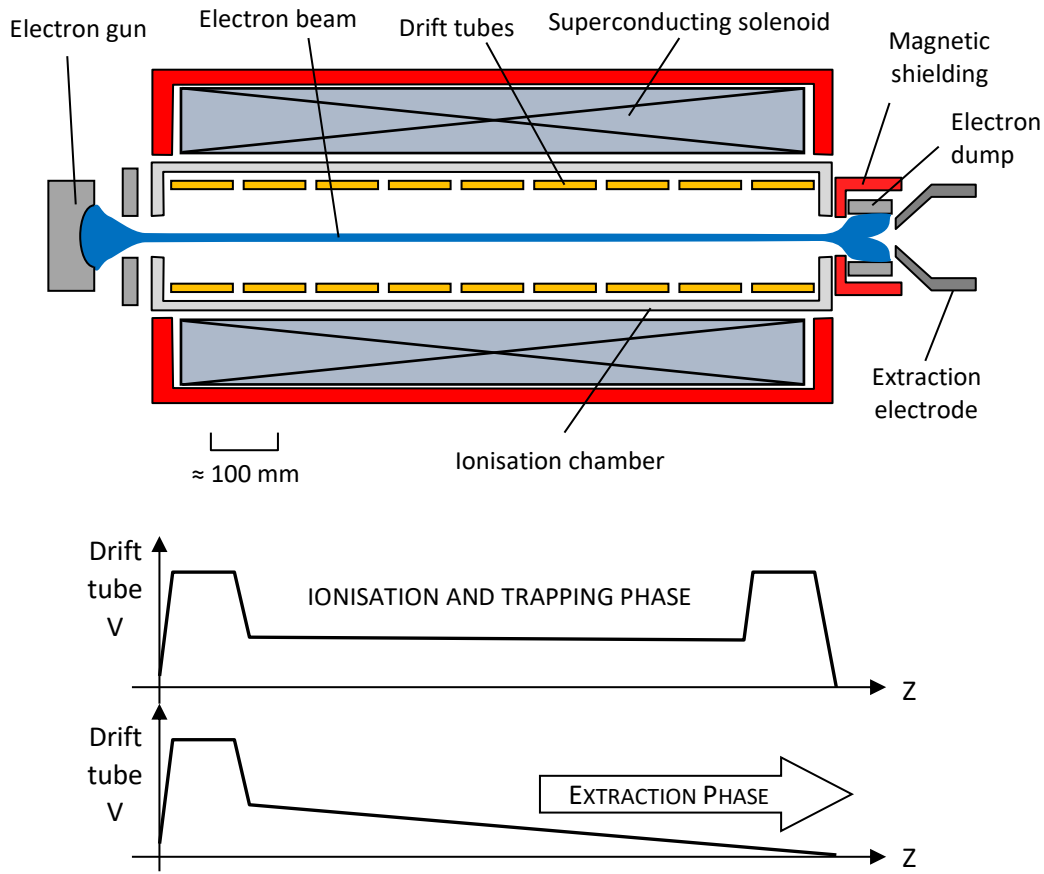

**Fig. 22:** Schematic of an electron beam ion source and drift tube voltage distrbutions for ionisation and extraction phases.

The material to be ionised is either pulsed into the middle of the ionisation chamber or injected as a low energy, low charge state beam from another ion source. The strong space charge of the negative electron beam creates a potential well that traps the injected positive ions. The amount of charge that can be trapped is limited by the size of the potential well created by the electron beam. Once trapped the ions undergo successive ionisations by the electron beam. During the trapping and ionisation phase higher positive voltages are applied to the end drift tubes to longitudinally confine the ions as shown in Fig. 22. Once the required charge state has been reached the extraction phase begins by modifying the potential distribution on the drift tubes as shown in Fig 22. The EBIS has been developed by many researchers, most recently by Jim Alesi and his team at BNL to produce a 1.7 emA, 10 μs, 5 Hz beam of $Ag^{32+}$ ions.

Due to the sequential injection, ionisation, extraction phases it is only possible to run an EBIS source in pulsed mode. Precise control of the ionisation phase in an EBIS produces a beam with a narrower charge state distribution compared to an ECR source, but downstream m/q selection in a dipole is still necessary if a single charge state beam is required. If the ionisation time in an EBIS is extended the source can be described as an Electron Beam Ion Trap (EBIT).

Like ECR, EBIS sources are used as 'charge breeders' in experiments with radioactive beams.

# 6 Negative Ion Sources

## 6.1 Introduction

### 6.1.1 The Negative Ion

Negative ion sources produce beams of atoms (or molecules) with an additional electron. The binding energy of the additional electron to an atom (or molecule) is termed the electron affinity. Some elements have a negative electron affinity (such as beryllium, nitrogen or the noble elements) which means they cannot form stable negative ions. H⁻ is a commonly used negative ion beam. Hydrogen has an electron affinity of 0.7542 eV. Considering the electron binding energy of neutral hydrogen is 13.6 eV the extra electron on an H⁻ ion is very loosely held on. All the ion sources in this section have been used to produce D⁻ ions as well as other heavy negative ions, such as O⁻, B⁻, C⁻ etc.

### 6.1.2 Uses

Negative ion sources were initially developed to allow electrostatic accelerators to increase their output beam energy. In an electrostatic accelerator, negative ions are first accelerated from an ion source at ground potential to a positive high voltage terminal. Inside the high voltage terminal the negative ions pass through a stripper (a very thin foil or a gas cell) which strips off electrons. The stripper changes the charge state of the beam from -1 to +1 (and higher). The resulting positive ions are then further accelerated from terminal volts back to ground. Accelerators using this technique are called tandem accelerators.

Cyclotrons use negative ions and stripping foils to extract the beam from the cyclotron. The stripping foil is positioned near the perimeter of the cyclotron poles. As the negative ion beam is accelerated it circulates on larger and larger radii until it passes through the stripping foil, which converts the beam from being negative to positive. The Lorenz force on the beam is reversed and instead of the force pointing into the centre of the cyclotron it points outwards and the beam is cleanly extracted.

In high power proton accelerators H⁻ ions allow charge accumulation via multi-turn injection. An H⁻ beam from a linear accelerator is fed through a stripping foil into a circular ring (a storage, accumulator or synchrotron ring) leaving protons circulating in the ring. The H⁻ beam from the linear accelerator continues to enter the ring whilst the circulating beam repeatedly passes through the stripping foil largely unaffected. The incoming beam curves in one way through a dipole as an H⁻ beam, then it curves out of the dipole in the opposite direction as a proton on top of the circulating beam. This allows accelerator designers to beat Liouville's theorem and build-up (paint) charges in phase space. Without this negative ion stripping trick it is very hard to accumulate more than one turn in a ring.

Fusion machines require very high current fast neutral beams of H or D atoms to be injected into the fusion plasma. The most efficient way of producing a neutral beam at 1 MeV is by first accelerating negative ions to 1 MeV then passing them through a gas cell neutraliser to strip them to neutrals.

## 6.2 Physics of Negative Ion Production

### 6.2.1 Mechanisms

The physical processes involved with the production of negative ions can be generally described as: surface, volume and charge exchange production processes. In different types of source one production process may dominate, however all three processes might contribute to the overall extracted negative ion current.

### 6.2.2 Charge Exchange

The first H⁻ ion sources were charge exchange devices. There are two ways of doing this: with foils or gas cells. With foils, a proton beam with an energy of about 10 keV is passed through a negatively biased foil and by electron capture an H⁻ beam is produced. With gas cells the proton beam is passed through a region filled with a gas and the H⁻ beam is produced by electron capture by dissociation of atomic or molecular ions in the gas. With foils or gas cells, only about 2% of the protons are converted into H⁻ ions. Until the 1960s this was the main technique used to make H⁻ beams. Beams of up to 200 μA were produced using this method.

Resonant charge exchange between fast H⁻ ions and slow neutral hydrogen atoms is essential to the operation of a negative Penning source.

### 6.2.3   Surface Production

For several decades numerous researchers [15, 16] had been experimenting with sources originally designed to produce positive ions, by reversing the polarity of the extraction they were able to extract negative ions. However, the co-extracted electron current was always at least an order of magnitude higher than the negative ion current.

In the early 1960s George Lawrence and his team at Los Alamos [17] were using a duoplasmatron to produce H⁻ ions when they first noticed that substantially higher beam currents and lower electron currents could be extracted when the extraction was off-centred (displaced) from the intermediate electrode (Fig. 23). They concluded that the extracted H⁻ ions must be being produced on the electrode surface or near the edge of the plasma. (This was also discovered independently by a team at the UK Atomic Weapons Establishment [18]).

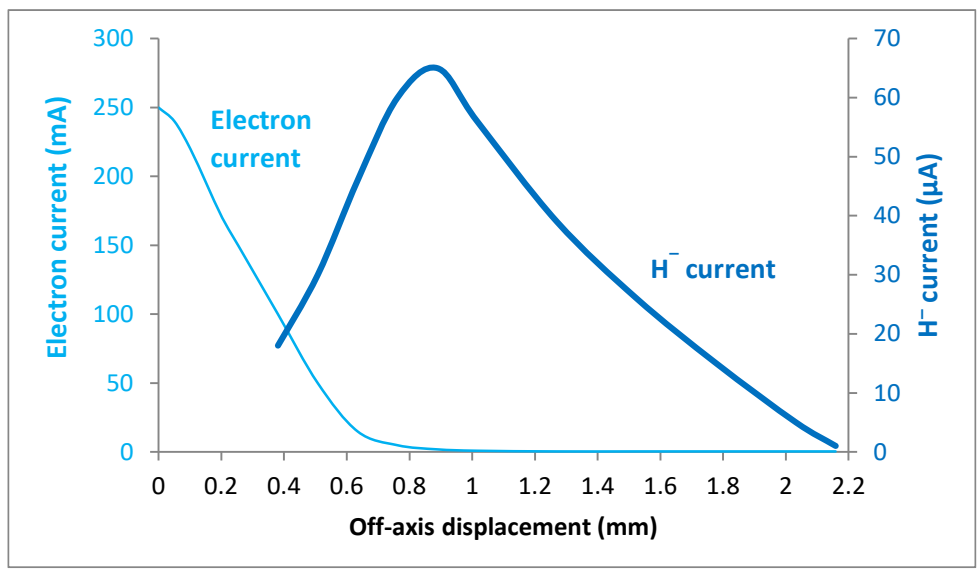

**Fig. 23:** H⁻ and electron currents as a function of extraction offset in a duoplasmatron measured at Los Alamos National Laboratory.

During the 1960s various sources originally designed to produce positive ions were adapted and modified to produce H⁻ ions and beam currents up to a few mA were produced.

### 6.2.4   Caesium Effect

In the early 1960s Victor Krohn, Jr. at Space Technology Laboratories, Inc. California was experimenting with surface sputter ion sources [19]. Surface sputter ion sources are used to produce beams of heaver ions (such as metals) for coating and etching applications. Krohn noticed that when $Cs^+$ ions were used to sputter a metal target the yield of sputtered negative ions increased by an order of magnitude. Caesium sputter sources (see Section 6.3) are used to produce low currents of many different negative ion species for tandem accelerators.

In the early 1970s Dimov and the team at the Budker Institute of Nuclear Physics tried adding caesium vapour to a magnetron source and achieved a record breaking 22 mA H⁻ beam current. This success led another member of the Budker team, Vadim Dudnikov, to develop a Penning type source that could produce 150 mA of H⁻ beam current with only 250 mA of extracted electrons. These type of caesiated cold cathode negative ion sources were named Surface Plasma Sources (SPS) (see Section 6.5) by the Budker researchers. When these results were published interest in caesiated sources took off and researchers all over the world started using caesium in their ion sources and a large number of new negative ion source designs were developed.

A very different type of H⁻ source that relies on surface production of H⁻ ions is the Surface Converter Source (see Section 6.4.3). Developed in the 1980s by Ken Ehlers and Ka-Ngo Leung at the Lawrence Berkeley Laboratory, it also relies on a caesiated surface. The caesiated surface sits in the middle of the plasma and is curved with a radius centred on the extraction region. H⁻ ions produced on this surface are 'focused' towards the extraction hole because they are repelled by the negative potential on the converter surface, this is why this type of source is also called 'self extracting'. H⁻ beam currents of greater than 1 A have been produced by this type of source.

### 6.2.5 *Surface Physics Processes*

The most important factor affecting H⁻ production at a surface is the work function, W. To make H⁻ ions the surface must provide electrons, so a low work function surface is essential. The work function of a surface obviously depends on what it is made of. If different atoms of a different element are adsorbed on that surface (such as caesium) then the work function can be altered. The 'thickness' of the adsorbed layer has have an effect on the surface's work function.

The thickness of the adsorbed layer is usually defined in terms of the number of 'monoloayers' of the adsorbed atoms.

$$Thickness\ (number\ of\ monolayers) = \frac{Number\ of\ adsorbate\ atoms\ per\ unit\ area}{Number\ of\ adsorbate\ atoms\ for\ a\ monolayer\ per\ unit\ area} \quad (17)$$

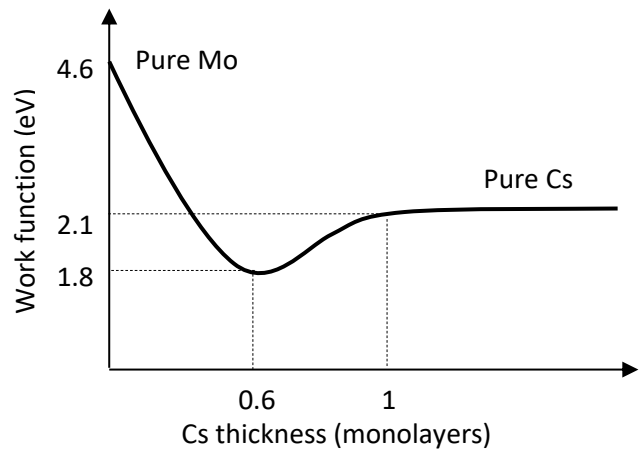

**Fig. 24:** Surface work function vs caesium thickness on a molybdenum surface.

When talking about negative ion production the surface is usually the cathode and is typically made of a high melting point metal such as tungsten W = 4.55 eV or molybdenum W = 4.6 eV. Caesium has the lowest work function of all elements: W = 2.14 eV. The work function of a caesium coated surface is actually lower than that of bulk caesium. As caesium covers the surface the work function decreases to 1.8 eV at 0.6 of a monolayer and then rises to about 2 eV for one monolayer or greater of caesium as shown in Fig. 24. The Cs-Cs bond is quite weak, so above room temperature no more than a monolayer will ever be deposited on a surface.

*6.2.6    Maintaining Caesium Coverage*

To minimise the work function and maximise the H⁻ production an optimum layer of caesium must be maintained on the surface. The surface is a dynamic place, caesium atoms are constantly being desorbed by plasma bombardment. To maintain optimum caesium coverage a constant flux of caesium is required. This is often provided by an oven containing pure elemental caesium but it could also be provided by heating locally positioned caesium chromate cartridges.

For pure caesium, the flux of caesium into the plasma can be precisely controlled by setting the temperature of the caesium oven. Figure 25 shows how the vapour pressure of caesium varies with temperature [20]. Sources that use elemental caesium in an oven generally operate between 140°C and 190°C.

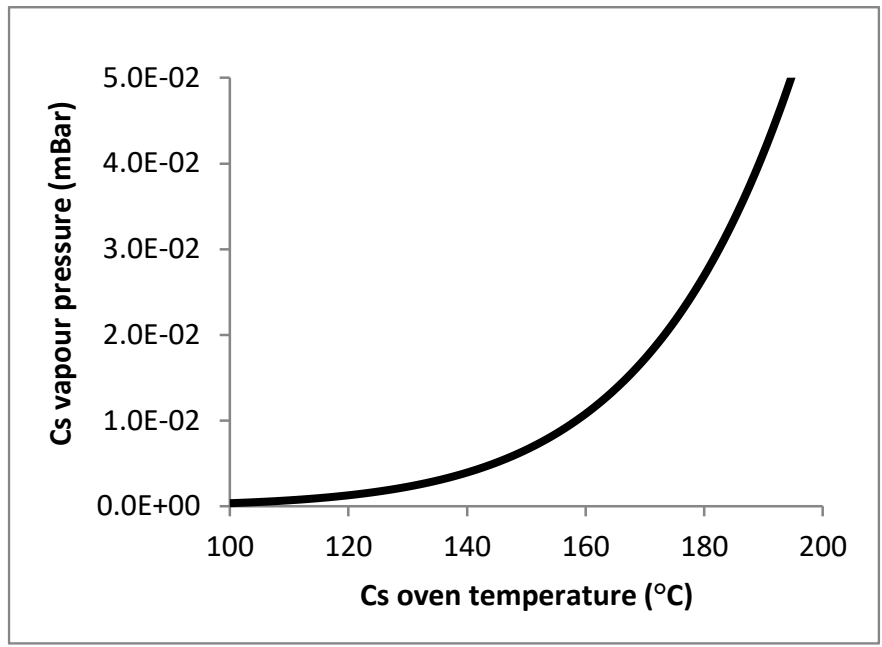

**Fig. 25:** Caesium vapour pressure vs caesium oven temperature.

In addition to aiding H⁻ surface production, caesium also helps to stabilise the plasma by readily ionising to produce additional electrons for the discharge.

*6.2.7    Volume Production*

Also, in the 1970s in parallel to the discovery of caesium enhanced H⁻ surface production, Marthe Bacal at École Polytechnique developed a completely new type of source that relied on H⁻ production in the plasma volume itself. Initially people were sceptical because H⁻ ions are so fragile: only 0.7542 eV is required to detach the extra electron. The plasma in the discharge was thought to be too energetic for any H⁻ ions produced in the volume to survive long enough to make it to the extraction region. The breakthrough that made volume production possible was the addition of a magnetic filter field near the extraction region. Bacal used a filament multicusp ion source (see Section 6.4.1) with a magnetic dipole filter field positioned near the extraction region. The magnetic filter field blocked high energy electrons from entering the extraction region, whereas ions and cold electrons could diffuse across the filter field. This effectively separated the discharge into two regions: a high temperature driver plasma on the filament side of the filter field, and a low temperature H⁻ production plasma on the extraction region side. Magnetically filtered multicusp ion sources are sometimes called 'tandem' sources because of these two regions of different plasma temperatures (not to be confused with tandem accelerators). The volume production process relies on the dissociative attachment of low energy electrons to rovibrationally excited $H_2$ molecules: $H_2^* + e\ (\leq 1\ eV) \rightarrow H^- + H^0$.

If the $H_2$ molecule is vibrationally cold the dissociative attachment cross section is extremely low ($10^{-21}$ cm$^2$). However, when the $H_2$ molecule is rovibrationally excited the cross section increases by 5 orders of magnitude. Thus, low energy electrons can be very effective in generating H$^-$ ions by dissociative attachment to rovibrationally excited $H_2$ molecules. The rovibrationally excited $H_2$ molecules are produced in the driver plasma region through inelastic electron collisions, then they drift through the filter field into the low electron temperature region. Rovibrationally excited molecules are also produced on the walls of the chamber and the electrode surfaces. A large proportion of the H$^-$ ions that are extracted are most likely to have been formed from rovibrationally excited molecules produced on the plasma electrode nearest the extraction hole.

### *6.2.8  H$^-$ Destruction*

In both volume and surface sources there are many processes that can destroy H$^-$ ions. The most common ones are:

$$H^- + H^+ \rightarrow H^0 + H^0 \quad \text{- Mutual Neutralisation}$$
$$H^- + e \rightarrow H^0 + 2e \quad \text{- Electron Detachment}$$
$$H^- + H^0 \rightarrow H_2^* + e \quad \text{- Associative Detachment.}$$

The aim of the source designer is to minimise the H$^-$ destruction processes by controlling the geometry, temperature, pressure and fields in the source. The following sections describe the design of the main negative ion sources used in particle accelerators.

## 6.3  Caesium Sputter Source

The caesium sputter source was developed by J.H. Billen and H.T. Richards at the University of Wisconsin and independently by Roy Middleton at the University of Pennsylvania in the late 1970s and early 1980s.

It is the development of the surface sputter ion source optimised for negative ion production using caesium. Sometimes called a 'Middleton source', it is sold commercially by National Electrostatics Corporation (NEC) as the SNICS (Source of Negative Ions by Caesium Sputtering). It can reliably produce low DC currents (10's of μA) of all negative ions. It is ubiquitously used in tandem accelerators.

The basic design of a caesium sputter source is shown in Fig. 26. The source runs at low pressures of about $2\times10^{-7}$ mBar. Caesium vapour is fed into the source from an oven and ionised on the hot surface of a conical or hemispherical electrode which is heated to around 1100°C. The Cs$^+$ ions are accelerated towards the cathode which is held at around -10 kV. The centre of the cathode is made of an insert material specified to produce the desired type of negative ions. A gas may also be introduced to the cathode insert region. After about 10-20 minutes of operation a sputter crater plasma forms in this region. The negative ions produced on the cathode are repelled by the negative bias voltage and extracted by a +15 kV extraction voltage applied to the extraction electrode. A magnetic field is used to supress the electrons that are also emitted from the cathode.

Middleton wrote a 'Negative Ion Cookbook' [21] containing recipes for making many negative ions including negative ion molecules using different combinations of cathode materials and gases. By using hydride, oxide, nitride or carbide molecules, elements that on their own cannot form negative ions can be accelerated in tandem accelerators. When a 2 atom negative molecule goes through the stripping foil it disintegrates into 2 positive ions.

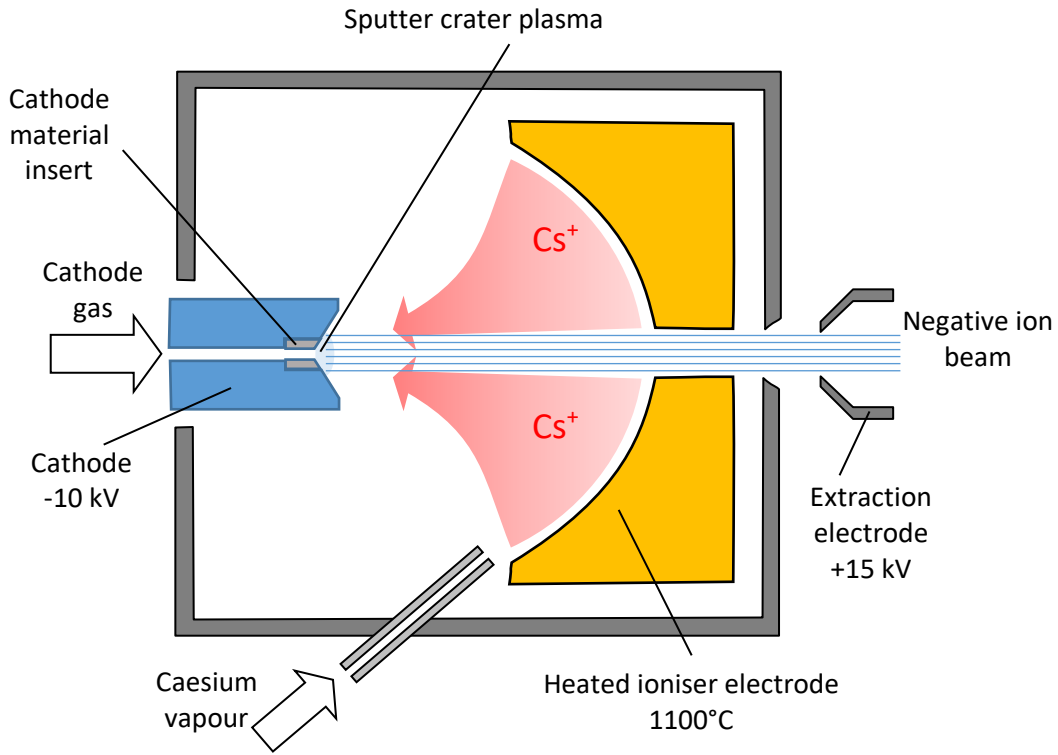

**Fig. 26:** Schematic of a cesium sputter negative ion source.

## 6.4 Hot Cathode Negative Ion Sources

### 6.4.1 Filament Cathode Volume Source

The filament cathode volume source was first developed by Martha Bacal at École Polytechnique, and by Ka-Ngo Leung and Kenneth Ehlers, at Lawrence Berkeley Laboratory in the 1970s. A schematic of the source is shown in Fig. 27. The filter field creates a low temperature plasma region that is conducive to H$^-$ production in the volume just in front of the extraction region. The filter dipole field can be created by a dedicated pair of magnets as shown in Fig. 27 or by modifying the multicusp arrangement around the edge of the plasma chamber.

This type of source has been successfully developed all over the world and a few companies now sell them commercially. They are reliable and although they only have a lifetime of a few weeks they are very low maintenance, only requiring a very simple filament change. They are used mainly on cyclotrons.

Andrew Holmes and the team at Culham found that the maximum current that can be extracted from this type of single aperture source is about 40 mA DC, it is limited by the maximum H$^-$ density obtainable at the extraction region. The H$^-$ current does not significantly increase above discharge currents of about 200 A because H$^-$ destruction processes start to dominate.

If caesium is added to this type of source the H$^-$ current can be doubled to 80 mA. The caesium does not increase the volume production of H$^-$, instead it facilitates surface production making this source a combined volume and surface source.

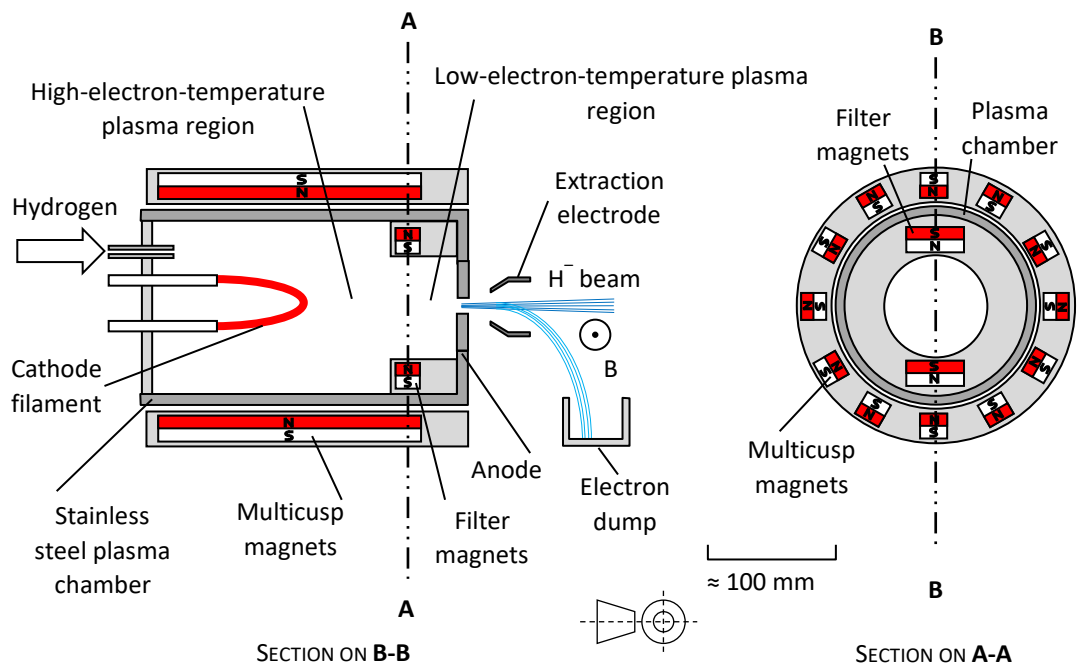

**Fig. 27:** Schematic of a filament cathode volume source.

### *6.4.2 LaB$_6$ Cathode Volume Source*

The hot cathode does not have to be made of tungsten wire. The first source used at J-PARC employed a coil made from LaB$_6$ as a cathode. The LaB$_6$ cathode is heated to 1500°C so that it thermionically emits electrons. A 20 mA at 1.5 % duty factor (600 μs and 25 Hz) beam was produced. By adding caesium the output current can be doubled.

### *6.4.3 Filament Cathode Surface Converter Source*

The surface converter source consists of a large multicusp filament discharge as shown in Fig. 28. The H$^-$ ions are produced on a caesiated molybdenum converter surface opposite the extraction hole. It was first developed by Ehlers and Leung at Lawrence Berkeley University in the late 1970s and early 1980s.

Caesium vapour is fed into the discharge chamber and it coats the converter electrode surface. This type of source is sometimes called a self-extracting source because: 1) the converter is negatively biased so H$^-$ ions produced on its caesium coated surface are repelled toward the extraction hole, and 2) The radius of curvature of the converter surface is centred on the extraction hole to focus the H$^-$ ions towards the extraction aperture. The multicusp magnets on either side of the extraction region act as a filter field to allow some volume production of H$^-$ ions to supplement the surface converter produced ions. A discharge of several hundred amps is created between the filament cathode and the anode walls. The multicusp field (as shown in Fig. 28) on all the walls confines the plasma.

This type of source is used on the LANSCE accelerator at Los Alamos National Laboratory. It has been developed to produce DC beams up to 1 A by adding more cathode filaments (up to 6) and increasing the discharge current to 1000 A. This type of source has also been used to generate other negative ions.

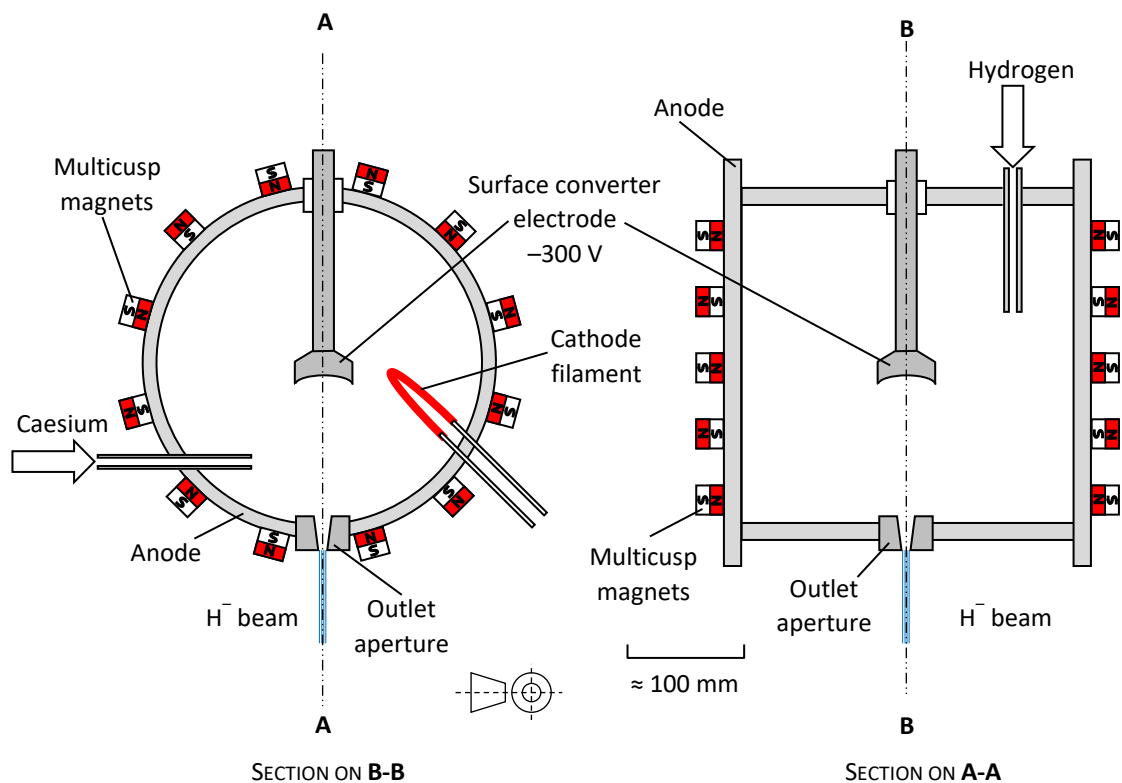

**Fig. 28:** Sectional schematic of a filament cathode surface converter source.

Like all filament discharge sources it suffers from lifetime limitations of a few weeks due to filament erosion. This source takes about 10 hours to start up and stabilise its output. It takes this long to develop an equilibrium coverage of caesium on the surface converter. Long start up times can impose operational restrictions on the rest of the machine.

## 6.5 Cold Cathode Negative Ion Sources – Surface Plasma Sources (SPS)

### 6.5.1 *Magnetron Surface Plasma Source*

The negative magnetron surface plasma source has a racetrack shaped discharge bounded on the inside by the cathode and the outside by the anode as shown in Fig. 29. The anode and cathode are only about 1 mm apart so the discharge is in the shape of a ribbon wrapped around the cathode. A magnetic field of between 0.1 to 0.2 T is applied perpendicular to the plane of the racetrack, this causes the plasma to drift around the racetrack. On one of the long sides of the racetrack discharge the anode has a hole through which the beam is extracted. Pulsed hydrogen is fed into discharge on the opposite side to the extraction hole. Caesium vapour is introduced via an inlet on one side.

The magnetron was the first ion source where the H⁻ current was first significantly increased by adding caesium vapour. This work was done by Gennadiĭ Dimov, Yuri Belchenko and Vadim Dudnikov at the Budker Institute of Nuclear Physics in the early 1970s. Magnetron sources have been developed by: Krsto Prelec and Jim Alesi at Brookhaven National Lab (BNL); by Chuck Schmidt, Doug Moehs and Dan Bollinger at Fermilab; and by Jens Peters at DESY; and others.

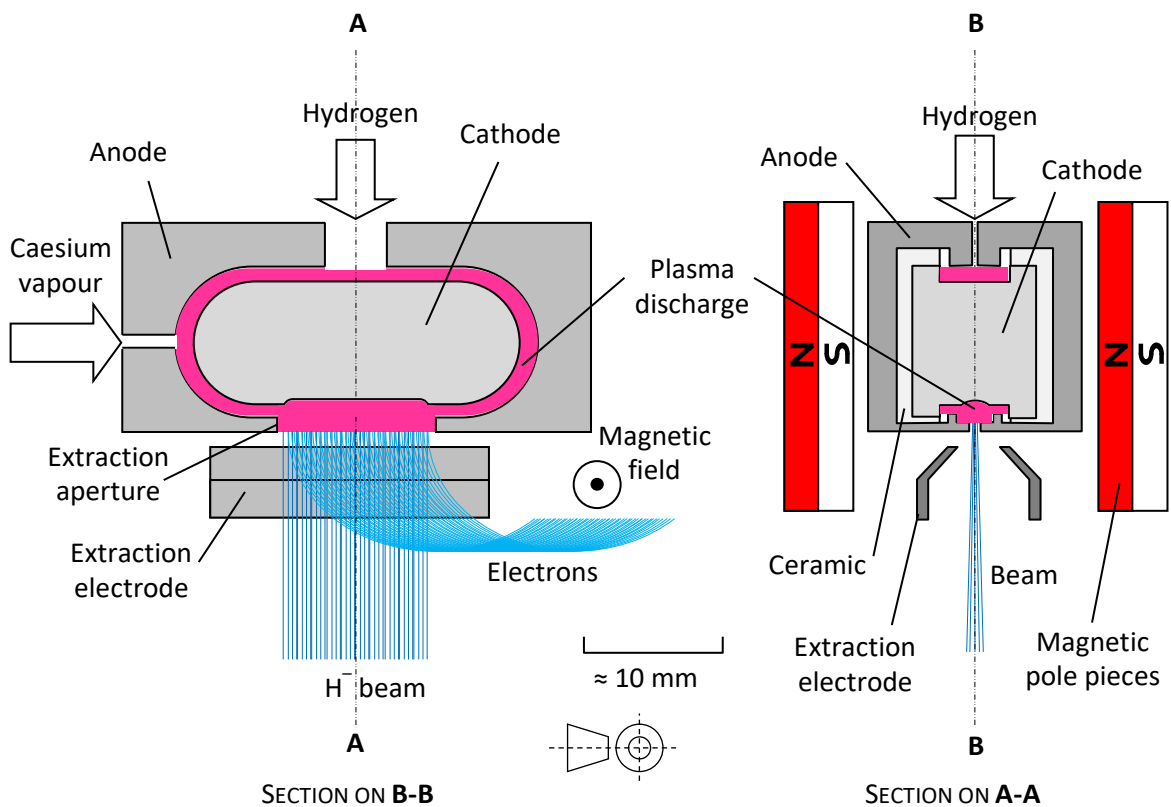

**Fig. 29:** Sectional schematic of a magnetron SPS with slit extraction.

A concave region in the cathode opposite the extraction hole gives a larger H⁻ production area and an initial focus to the extracted H⁻ beam. The magnetron can deliver high H⁻ currents of around 80 mA, but these currents have only been demonstrated at very low duty factors of up to 0.5% using the design shown in Fig. 29. The cathode does not have adequate cooling for higher duty cycles.

BNL developed a large multi-aperture magnetron that produced a 2 A H⁻ beam for fusion neutral beam injectors. For similar reasons the Soviets built a huge 11 A H⁻ beam current device called a Semi-planotron, which is just half of the magnetron racetrack, the discharge does not do a complete lap of the cathode.

### *6.5.2  Penning Surface Plasma Source*

The Penning surface plasma source (shown in Fig. 30) used on the ISIS accelerator at the Rutherford Appleton Laboratory (RAL) has a small (10 mm × 5 mm × 5 mm) rectangular discharge region with a transverse magnetic field. The long sides of discharge are bounded by two cathodes, with the other 4 walls at anode potential, creating a 'window frame' anode that completes the Penning discharge geometry. The magnetic field is orientated orthogonally to the cathode surfaces so that electrons emitted from the cathodes reflex back and forth between the parallel cathode surfaces, confined by being forced to spiral around the magnetic field lines. The primary anode is hollow and has holes through which hydrogen and caesium vapour are fed into the discharge. The plasma electrode aperture plate is also at anode potential. The beam is extracted from the plasma through a slit in the aperture plate by a high voltage applied to the extraction electrode. The electrodes are made of molybdenum.

H⁻ ions are produced on the cathode surfaces and accelerated by the plasma sheath fields that exist next to the cathodes. The plasma sheath potential is about 60 V. This produces fast H⁻ ions, but the geometry of the Penning source means that these fast cathode produced H⁻ ions cannot directly reach the aperture in the plasma electrode. The fast H⁻ ions undergo resonant charge exchange with slow neutral hydrogen atoms in the plasma producing slow H⁻ ions that drift towards the aperture in the plasma electrode where they are extracted in the H⁻ beam. This charge exchange process means the energy spread of the beam from a Penning SPS is much lower than from a magnetron SPS and the beam current is significantly less noisy.

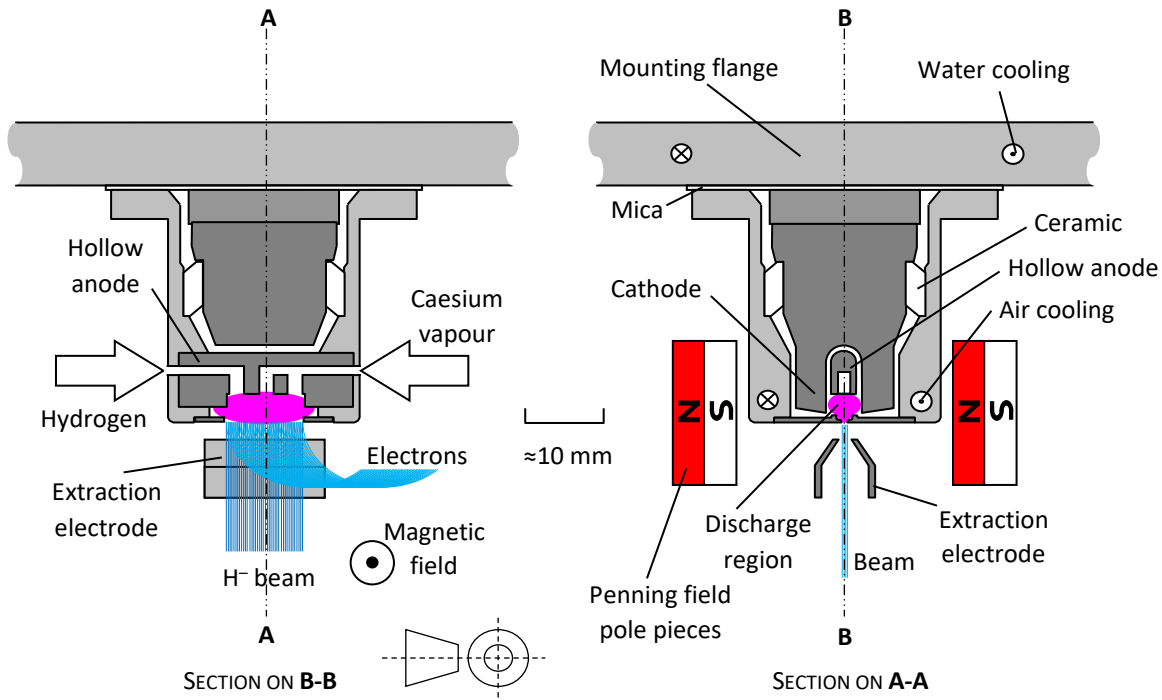

**Fig. 30:** Sectional schematic of a Penning SPS H⁻ ion source with slit extraction.

The H⁻ Penning ion source was first developed by Vadim Dudnikov and the Budker team in the early 1970s. Dudnikov demonstrated high pulsed H⁻ currents up to 150 mA and duty cycles all the way up to DC. Joe Sherman, Vernon Smith, Paul Alison and the team at Los Alamos developed scaled up Penning sources that also gave very high currents with low emittances. Dan Faircloth and the team at RAL have also developed a 2X scaled version of the operational ISIS Penning source that can produce 70 mA 2 ms 50 Hz H⁻ currents and up to 150 mA at shorter pulse lengths [22].

Yuri Belchenko and the Budker team have developed a DC Penning SPS capable of producing a 20 mA DC H⁻ beam current [23].

The Penning SPS is the brightest H⁻ ion source with current densities at extraction well above 1 Acm$^{-2}$ easily achievable.

Although classified as 'cold cathode' sources, the electrodes in both the Penning and magnetron SPS are heated by the discharge and run optimally between 200-500°C. Depending on the duty cycle additional cooling may be required.

The lifetime of the Penning source (when run at high duty cycles) is limited to a few weeks because of cathode sputtering by caesium ions. Both magnetron and Penning negative ion sources are fundamentally lifetime limited by sputtering erosion and material redistribution processes that lead to blockages and short circuits. When run at high currents they both have a lifetime of around 1 discharge day i.e. a DC source can run for a total of 24 hours before requiring refurbishment, whereas a source that runs with a 1% duty factor could keep going for 100 days.

## 6.6 RF Negative Ion Sources

### 6.6.1 Introduction

Attempts have been made to produce negative ions at all frequencies, but the most successful have been with inductively coupled plasmas in the low MHz range. This is because. ECR sources in the GHz range produce higher energy electrons that are not useful for H$^-$ ion production.

### 6.6.2 Internal RF Solenoid Coil Negative Ion Source

Having the coil immersed in the plasma provides good coupling, but the coil must be covered by an electrically insulating coating. Figure 31 shows a schematic of the internal RF solenoid coil multicusp volume source which includes a filter field to screen the H$^-$ production region from fast electrons.

In the early 1990s Ka-Ngo Leung and the team at Lawrence Berkeley Laboratory developed this type of source with a two and a half turn coil inside the plasma chamber. The coil was made of 4.7 mm diameter copper tubing and coated with porcelain. It was powered using a 2 MHz 50 kW RF amplifier. The coil was water-cooled. They obtained an H$^-$ current of about 40 mA which could be increased to about 90 mA by adding caesium. The lifetime is limited to a few weeks due to coil erosion.

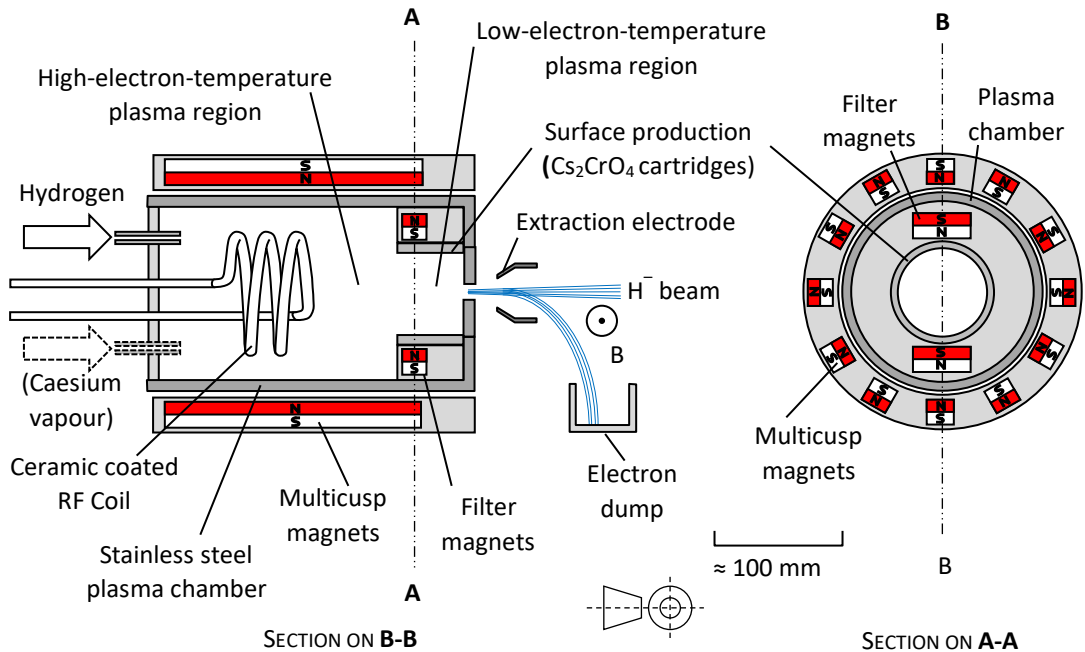

**Fig. 31:** Schematic of an RF solenoid internal coil multicusp negative ion source. Without caesium the source runs in pure volume production mode. Caesium can be added as a vapour from an external oven or by heating caesium chromate Cs$_2$CrO$_4$ cartridges.

This type of source is currently being used operationally for SNS at Oakridge National Laboratory with heated caesium chromate cartridges employed to increase H$^-$ surface production at the extraction region. 50 mA H$^-$ beams at 60 Hz 6% duty factor are routinely produced. This source technology is also used at J-PARC but with elemental Cs in an oven.

### 6.6.3 External RF Solenoid Coil Negative Ion Source

The problem of coil erosion in multicusp sources is avoided by putting the coil outside the plasma chamber as shown in Fig. 32 Jens Peters and his team at DESY were the first laboratory to successfully try this in the late 1990s. They put 50 kW into a three turn solenoid coil outside an Al$_2$O$_3$ ceramic chamber and obtained a 40 mA H⁻ beam with a duty factor of 0.05% and a pulse length of 100 µs without caesium. The source ran for a record breaking 300 days. They also experimented with different numbers of turns on the coil and different frequencies ranging from 1.65 to 9 MHz. The optimum appeared to be about 5 turns and 2 MHz.

Figure 32 shows a 5 turn external solenoid coil multicusp source. The pulsed hydrogen must be pre-ionised by a spark gap (or other method) when it is injected to the plasma chamber. A filter field is positioned in front of the extraction aperture.

Development work is on-going at RAL to increase the duty factor of this type of source to 1% at 50 Hz with a 30 mA H⁻ beam current.

Caesium can be added to external coil sources to double the H⁻ beam current, but problems associated with caesium result in shorter lifetimes.

CERN have just installed a caesiated external coil source with the new LINAC4 using elemental caesium fed from an oven and SNS are developing an external coil source using caesium chromate cartridges.

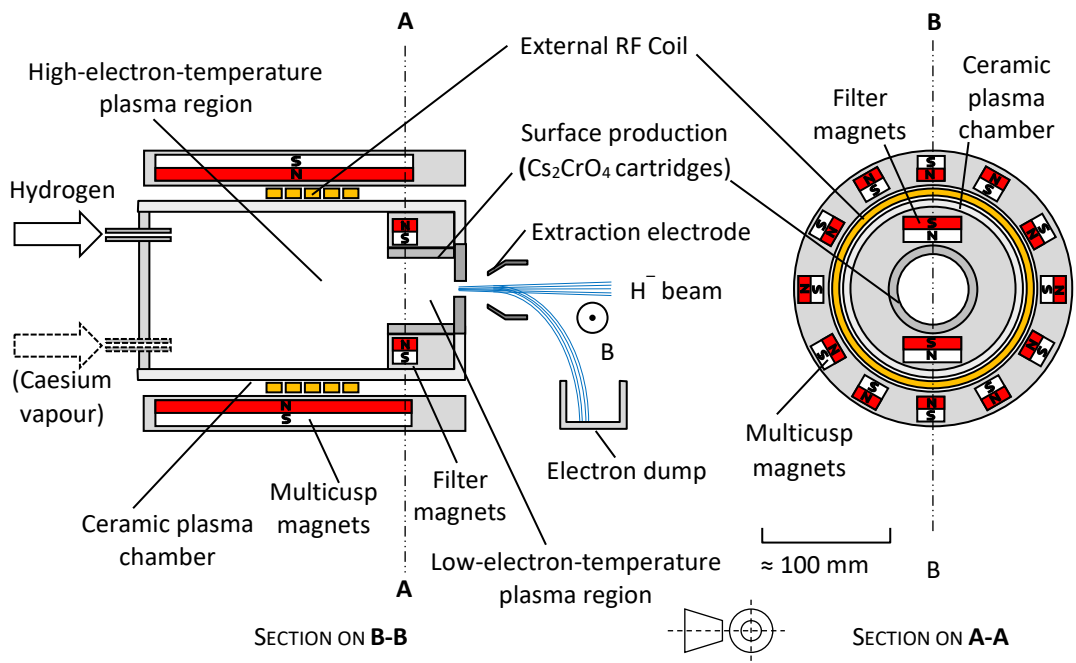

**Fig. 32:** Schematic of an external RF solenoid coil multicusp negative ion source. Without caesium the source runs in pure volume production mode. Caesium can be added as a vapour from an external oven or by heating caesium chromate Cs$_2$CrO$_4$ cartridges.

### *6.6.4   External Planar RF Coil Volume Source*

The RF coil does not necessarily have to be solenoidal. In an external planar RF coil volume source the plasma discharge is coupled to a flat spiral RF coil behind a dielectric RF window as shown in Fig. 33. The plasma discharge is cylindrical with a magnetic dipole filter field near the emission aperture. Hydrogen is fed into the discharge through a hole. The walls of the vessel are lined with a multicusp magnetic field arrangement to confine the plasma.

This source was developed by D-Pace and the University of Jyväskylä, Finland [24]. This source is commercially available to purchase from D-Pace as a complete 'turnkey' system. With a 3 kW 13.56 MHz RF discharge, this source can produce up to 8 mA CW H⁻ beams. A simple aluminium-nitride RF window (shown in Fig. 33) is the only relatively-fragile plasma-facing part; so long lifetimes can be achieved. This source can also produce D⁻ and $C_2^-$ beams.

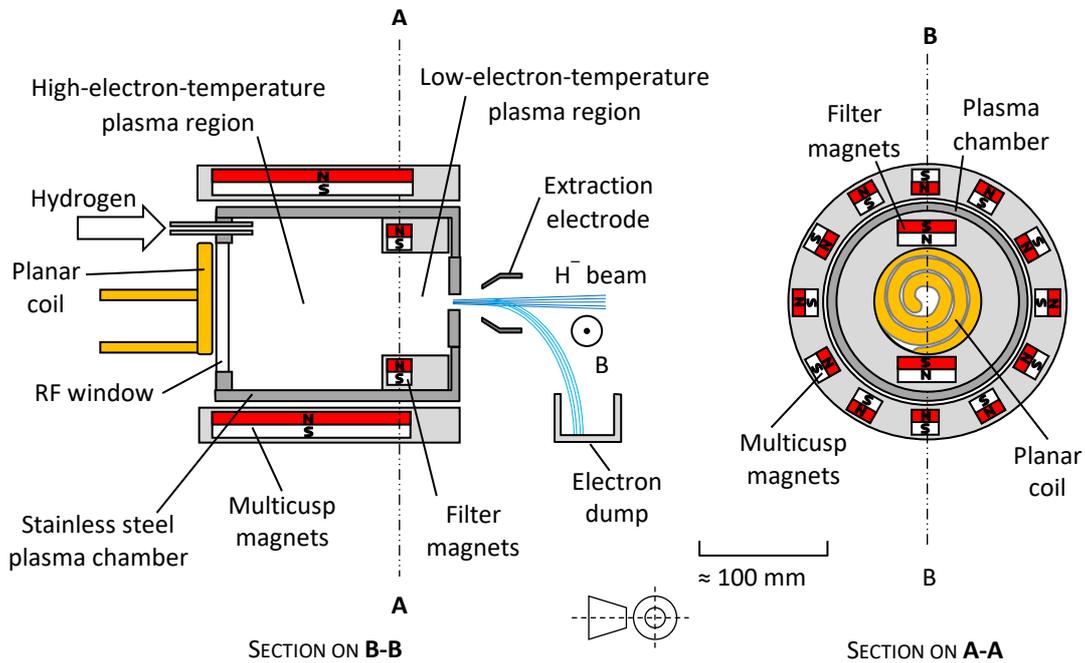

**Fig. 33:** Schematic of an external planar RF coil volume source.

## 7 Polarised Sources

### 7.1 Polarised Electron Guns

Polarised electron beams are created using strained gallium arsenide photocathodes illuminated with a circularly polarised laser beam. The Jefferson Lab photocathode gun can produce a 100 µA polarised electron beam with an 80% polarisation fraction.

### 7.2 Polarised H⁻ Sources

Using a multistage process it is possible to produce beams of polarised H⁻ beams. Anatoli Zelenski and the team at BNL developed an Optically Pumped Polarized H⁻ Ion Source (OPPIS) [25]. The complicated system starts with a high current proton source feeding beam into a hydrogen neutraliser cell. The neutralised beam then passes into a helium ioniser cell, then it enters a laser pumped rubidium-vapour cell before going through a sona-transition before finally being ionised in a sodium jet cell. 1.6 mA 400 µs polarised H⁻ beams are routinely produced for RHIC.

Polarised sources are a huge and fascinating subject (see Zelenski's review [26]), but unfortunately (like for all the other sources in this paper!) there is no time to go into any detail.

## 8 Running and Developing and Sources

### 8.1 Which Source?

The type of source an accelerator uses obviously depends on what type of particles are required and what beam current is needed. However, the reason why one type of source is used and not another is usually historic. It depends on when the accelerator was built, what facility was there previously and what expertise is available. Cost can sometimes be an issue as well, especially for the highest beam currents. Filament, Penning and magnetron sources are the most affordable.

CERN used a duoplasmatron because it was the only source capable of producing a 500mA proton current. Fermilab use a magnetron SPS because it was the best source that met their needs when it was developed in the late 1970s. Similarly, RAL use a Penning SPS because it was the best source that met their needs in the early 1980s.

If you were building a new machine today which source would you choose?

For electron beams there are numerous gridded thermionic dispenser cathodes available from a number of manufactures. The cathode needs to be mounted in an accelerating structure (RF or DC) that meets the requirements of the rest of the machine. For very high pulsed current or polarised beam applications a photo-cathode is the best choice.

For proton sources up to 100 mA the microwave discharge ion source is the obvious choice because it offers exceptional lifetimes and reliability. For even higher proton currents of a few hundred mA the duoplasmatron or gas dynamic ECR sources are the only options available. For heavy or multiply charged ions the ECR ion source is the best option. For very high charge state ions the EBIS is the best option. For beams of elements that are solid the vacuum arc or laser ion sources are a good option.

For low currents of a huge range of negative ions the caesium sputter source is ideal.

Intense development work continues at all the major labs operating high current $H^-$ ion sources.

Un-caesiated pure volume production filament ion sources offer DC currents up to 40 mA but only with very high filament and discharge currents. A commercial 15 mA DC $H^-$ filament volume source is available. At high currents and duty factors the filament has a short lifetime of only a few weeks, however maintenance is very easy.

Un-caesiated pure volume external RF coil sources offer the promise of great reliability and long lifetimes (>1 year). A commercial 8 mA DC source is now available. Pulsed currents of 40 mA have been demonstrated operationally at very low duty factors (0.05%), but current development work is pushing the duty factor to 1%.

Caesiated internal coil RF sources have demonstrated very high pulsed $H^-$ currents in excess of 100 mA and delivered 50 mA beams for accelerator operations at 6% duty factors, however in both cases lifetime is limited to a few weeks due to the insulating ceramic coating on the coil eroding away and due to the problems created by using caesium.

Surface converter sources have demonstrated high DC $H^-$ currents of 20 mA but only with short lifetimes of a few weeks and very long setup times (10 hours). Development sources have demonstrated 120 mA but with even shorter lifetimes.

Magnetron SPS have demonstrated high pulsed $H^-$ currents (80 mA) and very long lifetimes (>6 months) but only at very low duty factors (<0.5%).

Penning SPS have demonstrated high pulsed $H^-$ currents 60 mA (170 mA on experimental sources) and long duty cycles up to DC. Lifetime is limited to a few weeks at very high duty cycles.

## 8.2 Power Supplies and Control Systems

Ion sources present particular challenges for power supplies: they must deliver stable high currents to a plasma load that is often unstable; they must deliver stable high voltages to extraction electrodes that often breakdown. All the electrodes that the power supplies are connected to are often in close proximity and can be coated with caesium that increases the probability of sparking especially in an environment full charge carriers in the presence of strong magnetic fields. The power supplies must be very robust and stable. Overall system design is important because of the highly interconnected configuration. The entire ion source, power supplies and ancillary equipment are invariably installed on a high voltage platform so protection, grounding and isolation are essential details to get right. Digital control electronics are susceptible to errors in the event of inevitable high voltage breakdowns, so analogue control circuitry can be a more reliable option. Operation on a high voltage platform forces the need for some form of isolated control system to allow source tuning during operation and for the provision of timing signals. This is usually done over fibre optics, but some very old source designs use insulated mechanical linkages to change power supply settings.

## 8.3  Developing a Source

A lot of the ion sources and election guns discussed in this paper are still being developed by labs all over the world. Beam currents are being increased, duty cycles extended, emittances reduced and reliabilities improved. Modern finite element modelling and computational fluid dynamics allow the thermal, electrical and magnetic operation of the source to be investigated to a detail never before possible. Beam transport and plasma codes allow extraction and beam formation to be studied. All these computer modelling tools allow ion source development to progress without having to go through as many prototype iterations. However, it must be remembered that ion source behaviour is incredibly complex because it couples highly non-linear and often transient plasma phenomena with the control loop properties of numerous support systems. Ion sources often exhibit many emergent behaviours that could never be predicted by simulation alone. The only way to find out how a new source design will behave is to actually test it. This is why development rigs or test stands are essential to designing new sources. These test rigs should replicate the actual environment where the source will run, ideally it should also include a LEBT to test exactly what beam can be transported. With a LEBT included the test rig is more likely to accurately represent the vacuum pressure distribution, background gas ionisation, space charge compensation and back streaming particles that the ion source will see in operation. They must also run 24 hours a day if lifetimes are to be tested. Even then the true performance of the source will not be known until it runs on an operational machine for several years, exposed to variable conditions and inevitable human error.

A development test rig must be equipped with as many diagnostics as possible to try to understand how the source is performing. These could include:

- Beam current e.g. toroids, faraday cups
- Emittance e.g. slit-grid, pepperpot, slit-slit, Alison electric sweep scanner
- Profile e.g. scintillator, wire scanner, laser wire scanner
- Energy Spread e.g. retarding potential energy analyser
- Optical spectroscopy
- Langmuir probes.

## 9  Summary and Conclusions

Meeting the beam current, pulse length and emittance required by the accelerator is only part of the job of a particle source. Operational sources must be reliable and they must have a lifetime that is compatible with the operating schedule of the accelerator. They must be easy to maintain, and when they fail they must be easy to fix. If they have to be replaced they should be easy to dismantle. The start-up procedure should be made as quick and easy as possible.

Particle sources are a huge and fascinating subject. It can take a lifetime to become an expert in just one type of source. This paper has just provided an introduction to some of the most common particle sources in use today in particle accelerators.

For further reading see the Bibliography.

**Acknowledgements**

Many thanks to Olli Tarvainen for his careful reading of this paper and suggestions for improvement.

**References**

[1] F. Paschen  Wiedemann Ann. Phys. Chem. 37 69-96 (1889).

[2] A.J. Dempster, Phys. Rev. 8 (1916), 651-662.

[3] A.W. Hull, Phys. Rev. 18 (1921) 31–57.

[4] J.E. Brittain, Proc. IEEE 98 (2010) 635–637.

[5] L.R. Maxwell, Rev. Sci. Instrum. 2 (1931) 129.

[6] F.M. Penning, Physica IV(2) (1937) 71–75.

[7] P. Thonemann, Nature 158, 61 (1946).

[8] N. Sakudo, Rev. Sci. Instrum. 49, 940 (1978).

[9] J. Ishikawa, Y. Takeiri, and T. Takagi, Rev. Sci. Instrum. 55, 449 (1984).

[10] T. Taylor and J.F. Mouris, Nucl. Instrum. Methods 336 (1993).

[11] J. Sherman *et al*, Rev. Sci. Instrum. 69, 1003 (1998).

[12] R. Gobin, *et al,* Rev. Sci. Instrum. 73, 2, 922-924 (2002).

[13] O. Tuske, *et al*, Rev. Sci. Instrum. 79, 02B710 (2008).

[14] V. Skalyga, *et al,* Rev. Sci. Instrum. 85, 02A702 (2014).

[15] K.W. Ehlers *et al.*, Nucl. Instrum. & Methods, 22 (1963), 87-92.

[16] K.W. Ehlers, Nucl. Instrum. & Methods, 32 (1965), 309-316.

[17] G.P. Lawrence *et al.*, Nucl. Instrum. & Methods, 32 (1965), 357-359.

[18] L.E. Collins and R.H. Gobbett, Nucl. Instrum. & Methods, 35 (1965), 277-282.

[19] V.E. Krohn, Jr, Journal of Applied Physics 33, 3523 (1962).

[20] J.B. Taylor and I. Langmuir, Phys. Rev. 51 (1937), 753-760.

[21] https://www.pelletron.com/cookbook.pdf.

[22] D.C. Faircloth *et al*., AIP Conference Proceedings 2052, 050004 (2018).

[23] Y. Belchenko, *et al*., Review of Scientific Instruments 85, 02B108 (2014).

[24] T. Kalvas, *et al*., AIP Conference Proceedings 1655, 030015 (2015).

[25] A. Zelenski, *et al*., Rev. Sci. Instrum. 87, 02B705 (2016).

[26] A. Zelenski, Rev. Sci. Instrum. 81, 02B308 (2010).

**Appendix A: Bibliography**

**A.1  Papers**

M. Bacal, *Physics Basis and Future Trends for Negative Ion Sources*, Rev. Sci. Instrum. 79, 02A516, (2008).

M. Bacal, *Physics Aspects of Negative Ion Sources,* Nucl. Fusion 46 (2006) S250–S259.

L. Celona et al, *Microwave to plasma Coupling in Electron Cyclotron Resonance and Microwave Ion Sources,* Rev. Sci. Instrum. 81, 02A333 (2010).

S. Gammino et al, *Review on High Current 2.45 GHz Electron Cyclotron Resonance Sources,* Rev. Sci. Instrum. 81, 02B313 (2010).

C.E. Hill, *Ion and Electron Sources*, Proc. CERN Accelerator School, La Hulpe, Belgium, CERN-94-36 (1994).

T. Kalvas, *Beam Extraction and Transport,* Proc. CERN Accelerator School: Ion Sources, pp.537-564, (2013).

R. Keller, *High-intensity Ion Sources for Accelerators with Emphasis on $H^-$ Beam Formation and Transport*, Rev. Sci. Instrum. 81, 02B311, (2010).

F. Loehl, *High Current and High Brightness Electron Sources*, Proceedings of IPAC'10, Kyoto, Japan MOZRA01, (2010).

S. Nikiforov *et al.*, *Ion Sources for use in Research and Applied High Voltage Accelerators,* Proceedings of PAC95, 1004 - 1006 vol.2 (1995).

D.P. Moehs *et al.*, *Negative Hydrogen Ion Sources for Accelerators,* IEEE Trans. Plasma Sci. 33, 6, (2005), 1786-1798.

C.W. Schmidt, *Review of Negative Hydrogen Ion Sources,* Proceedings of LINAC 90, Albuquerque, NM, (1990).

J. Peters, *Review Of High Intensity $H^-$ Sources And Matching To High Power RFQ´S,* Proceedings of EPAC 2000, Vienna, Austria 113-117.

J. Peters, *Negative Ion Sources for High Energy Accelerators,* Rev. Sci. Instrum. 71, 2, (2000) 1069-1074.

C. W. Schmidt, *Historical Perspective of the $H^-$ Ion Source Symposia*, AIP Conf. Proc. Volume 439, 254-258 (1998).

R. Scrivens, *Electron and Ion Sources for Particle Accelerators,* Proc. CERN Accelerator School, Zeuthen, Germany, p494-504 (2003).

### A.2 Books

Ian G Brown, *The Physics and Technology of Ion Sources*, (Wiley-VCH, 2004).

Vadim Dudnikov, *Development and Applications of Negative Ion Sources*, (Springer Series on Atomic, Optical, and Plasma Physics, 2019).

Huashun Zhang, *Ion Sources*, (Science Press, 1999).

Bernhard Wolf, *Handbook of Ion Sources*, (CRC Press, 1995).

### A.3 Websites

Materials from the US Particle Accelerator Schools:

https://uspas.fnal.gov/materials/materials-table.shtml

Proceedings of the CERN Accelerator Schools:

https://cas.web.cern.ch/previous-schools